\documentclass[a4paper,fleqn,usenatbib]{mnras}

\usepackage[T1]{fontenc}
\usepackage{ae,aecompl}

\usepackage{graphicx}	
\usepackage{amsmath}	
\usepackage{amssymb}	
\usepackage{multicol}        
\usepackage{bm}		
\usepackage{pdflscape}	
\usepackage{hyperref}

\usepackage{float}

\title[BOSS DR12 combined galaxy sample]{The clustering of galaxies in the completed SDSS-III Baryon Oscillation Spectroscopic Survey: On the measurement of growth rate using galaxy correlation functions}

\author[S. Satpathy et al.]{
Siddharth Satpathy$^{1,2}$\thanks{E-mail: siddharthsatpathy@cmu.edu},
Shadab Alam$^{1,2}$,
Shirley Ho$^{1,2,3,4}$,
Martin White$^{3,4}$,
\and
Neta A. Bahcall$^{5}$,
Florian Beutler$^{6}$,
Joel R. Brownstein$^{7}$,
Chia-Hsun Chuang$^{8,9}$,
\and
Daniel J. Eisenstein$^{10}$,
Jan Niklas Grieb$^{11,12}$,
Francisco Kitaura$^{9,3,4}$,
Matthew D. 
\and
Olmstead$^{13}$,
Will J. Percival$^{6}$,
Salvador Salazar-Albornoz$^{11,12}$,
Ariel G. S\'{a}nchez$^{12}$,
\and
Hee-Jong Seo$^{14}$,
Daniel Thomas$^{6}$,
Jeremy L. Tinker$^{15}$   \& 
Rita Tojeiro$^{16}$
\\
$^{1}$Department of Physics, Carnegie Mellon University, 5000 Forbes Ave., Pittsburgh, PA 15213, USA\\
$^{2}$The McWilliams Center for Cosmology, Carnegie Mellon University, 5000 Forbes Ave., Pittsburgh, PA 15213, USA \\
$^{3}$Lawrence Berkeley National Laboratory, 1 Cyclotron Road, Berkeley, CA 94720, USA \\
$^{4}$Departments of Physics and Astronomy, University of California, Berkeley, CA, 94720, USA \\
$^{5}$Department of Astrophysical Sciences, Princeton University, Princeton, NJ, 08544, USA \\
$^{6}$Institute of Cosmology \& Gravitation, University of Portsmouth, Dennis Sciama Building, Portsmouth, PO1 3FX, UK\\
$^{7}$Department of Physics and Astronomy, University of Utah, 115 S 1400 E, Salt Lake City, UT 84112, USA \\
$^{8}$Departamento de F\'{i}sica Te\'{o}rica, Universidad Aut\'{o}noma de Madrid, Cantoblanco, 28049, Madrid, Spain\\
$^{9}$Leibniz-Institut f\"{u}r Astrophysik Potsdam (AIP), An der Sternwarte 16, D-14482 Potsdam, Germany\\
$^{10}$Harvard-Smithsonian Center for Astrophysics, 60 Garden St., Cambridge, MA 02138, USA\\
$^{11}$Universit\"{a}ts-Sternwarte M\"{u}nchen, Ludwig-Maximilians-Universit\"{a}t M\"{u}nchen, Scheinerstrasse 1, 81679 Munich, Germany\\
$^{12}$Max-Planck-Institut f\"{u}r extraterrestrische Physik, Postfach 1312, Giessenbachstr., 85741 Garching, Germany\\
$^{13}$Department of Chemistry and Physics, King's College, 133 North River St, Wilkes Barre, PA 18711, USA\\
$^{14}$Department of Physics and Astronomy, Ohio University, Clippinger Labs, Athens, OH 45701, USA \\ 
$^{15}$Center for Cosmology and Particle Physics \& Department of Physics, New York University, New York, NY 10003, USA \\
$^{16}$School of Physics \& Astronomy, University of St Andrews, North Haugh, St Andrews Fife, KY16 9SS, UK
}

\date{Accepted XXX. Received YYY; in original form ZZZ}

\pubyear{2016}

\begin{document}
\label{firstpage}
\pagerange{\pageref{firstpage}--\pageref{lastpage}}
\maketitle

\begin{abstract}
We present a measurement of the linear growth rate of structure, \textit{f} from the Sloan Digital Sky Survey III (SDSS III) Baryon Oscillation Spectroscopic Survey (BOSS) Data Release 12 (DR12) using Convolution Lagrangian Perturbation Theory (CLPT) with Gaussian Streaming Redshift-Space Distortions (GSRSD) to model the two point statistics of BOSS galaxies in DR12. The BOSS-DR12 dataset includes 1,198,006 massive galaxies spread over the redshift range $0.2 < z < 0.75$. These galaxy samples are categorized in three redshift bins. Using CLPT-GSRSD in our analysis of the combined sample of the three redshift bins, we report measurements of $f \sigma_8$ for the three redshift bins. We find $f \sigma_8 = 0.430 \pm 0.054$ at $z_{\rm eff} = 0.38$, $f \sigma_8  = 0.452 \pm 0.057$ at $z_{\rm eff} = 0.51$ and $f \sigma_8 = 0.457 \pm 0.052$ at $z_{\rm eff} = 0.61$. Our results are consistent with the predictions of Planck $\Lambda$CDM-GR. Our constraints on the growth rates of structure in the Universe at different redshifts serve as a useful probe, which can help distinguish between a model of the Universe based on dark energy and models based on modified theories of gravity. This paper is part of a set that analyses the final galaxy clustering dataset from BOSS. The measurements and likelihoods presented here are combined with others in \citet{Acacia2016} to produce the final cosmological constraints from BOSS. 
\end{abstract}



\section{Introduction}
The theory of General Relativity (GR) gives us a relation between the expansion rate of the Universe and its matter and energy content \citep[][]{Einstein1915, Einstein1916}. At the same time, cosmological observations have also given us a glimpse into the Universe's dark sector. The observation of the accelerated expansion of the Universe is a landmark discovery in cosmology \citep[][]{Riess1998, Perlmutter1999}. The acceleration of the expansion of the Universe is most commonly explained by a framework which suggests that our Universe is dominated by a `dark energy' field with negative pressure. The dark energy is similar to the cosmological constant ($\Lambda$) in Einstein's  theory of General Relativity \citep[][]{Padmanabhan2007}. The $\Lambda$CDM-GR model which proposed the accelerated expansion of the Universe is in consonance with probes such as the Cosmic Microwave Background (CMB) \citep[][]{Bennett2013, Planck2014a} and Baryon Acoustic Oscillations (BAO) \citep[][]{Eisenstein2005, Cole2005, Hutsi2006, Percival2007, Kazin2010, Percival2010, Reid2010}. The observation that the expansion of the Universe is accelerating can also pertain to the possibility of `dark gravity' \citep[][]{HenryCouannier2005a, HenryCouannier2005b, Bludman2007, Durrer2007, HenryCouannier2007, Heavens2009, Lobo2011, Lobo2012}, which suggests that General Relativity is incorrect on the largest scales and is a limit of a more complete theory of gravity. Such a possibility gives scope to the explanation of the accelerated expansion of the Universe by frameworks which try to reproduce cosmological observations by modifying the form of the equations of GR. The cause for the acceleration in the expansion of the Universe remains a mystery and one cannot settle on a preferred candidate to explain the measurement of the expansion history \textit{H(z)} from spectroscopic surveys and probes like Type Ia supernova and BAO since modified theories of gravity \citep[][]{Carroll2004, Kolb2006, Carroll2006, Cardone2012} and theories based on dark energy explain the observations equally well. In other words, the measurement of the expansion rate alone will not be able to distinguish between a model based on dark energy and modified theories of gravity. 
%
 
One way of resolving this conundrum lies in the investigation of the growth rate of structure inside the Universe \citep[][]{Peacock2006, Albrecht2007, Pouri2013, Pouri2014, Alam2015a, Mohammad2015}. The growth rate of structure in the Universe is decided by the competing effects of the gravitational collapse of density fluctuations, which accelerate their growth and the expansion rate, which inhibits it. Since the theory of General Relativity gives us a relation between the growth rate of cosmological structure and the expansion history of the Universe, measurements of the growth rate give us a handle on the underlying theory of gravity. In cosmological observations, the positions of galaxies are mapped by redshift, which correspond to the true distance according to the Hubble Law. Probes which look at the growth of structure in the Universe also include the peculiar velocities of galaxies. The observed redshift (\textit{z}) is, in fact, a sum of the Hubble recession velocity and the peculiar velocity caused by gravitational dynamics. Components from the peculiar velocities, which are deviations of galaxies' velocity from pure Hubble flow, combine with components from the Hubble flow to give rise to distortions in the reconstructed spatial distribution of the observed objects. The ensuing distortions manifest themselves as anisotropy in the distribution of objects and are caused in the radial direction in the redshift space \citep[][]{Kaiser1987, Hamilton1992, Cole1995, Guzzo1997, Peacock2001, Scoccimarro2004, Tegmark2006, White2009, Percival2009, Yoo2009, McDonaldP2009, McDonaldPSeljak2009, Percival2011, Reid2011, Yoo2012, Samushia2012, McQuinn2013, Beutler2014, White2015, Simpson2016}. These distortions are referred to as `redshift space distortions' (RSD).

Redshift space distortions can be used to reveal information about the motion of galaxies and underlying matter distribution in the Universe. The distinctive features of RSD are revealed in the two-point correlation statistics of galaxy distributions which are obtained as functions of variables representing distances parallel and perpendicular to the line of sight($s_{||}$ and $s_{\perp}$ respectively). In small spatial scales where galaxies with high velocities are dominant, RSD is manifested as elongation in redshift space maps with an axis of elongation pointing towards the observer (i.e., along $s_{||}$). This phenomenon is referred to as the ``Fingers of God'' effect \citep[][]{Jackson1972, Tegmark2004}. On larger scales, one observes the ``Kaiser effect'' where coherent peculiar velocities cause an apparent contraction of structure along the line of sight in redshift space. As a result, we see two distinct effects, $viz.$ the non-linear and the linear effects due to small-scale elongation and large scale flattening in redshift scale maps. Measurement of the growth rate of structure from RSD is intricate. In 1987, Nick Kaiser \citep[][]{Kaiser1987} tried to tackle this problem by introducing a prescription which discusses the redshift space power spectrum by modifying the linear theory of large scale structure. In his landmark work, Kaiser was able to relate the power spectrum in redshift space $P_s(\mathbf{k})$ and it's counterpart in real space $P_r(\mathbf{k})$  by the following relation:
\begin{align} \label{eqn:Kaiser}
P_s(\mathbf{k}) = \left( 1 + \beta \mu_k^2 \right)^2 P_r(\mathbf{k})
\end{align}
where $\mu_k$ corresponds to the cosine of the angle between $\mathbf{k}$ and the line of sight and $\beta = \Omega_m^{0.55}/b$ is the linear distortion parameter. Here, $\Omega_m$ is the mass density parameter and $b$ denotes the linear bias parameter. Exploration of the concept of peculiar velocities in non-linear scales using ideas of the ``streaming model'' was presented in \citet{Peebles1980, Davis1983, Fisher1995}. \cite{Peebles1980} showed that the factor $\Omega_m^{0.55}$ relates peculiar velocities to density fluctuations. The real space counterpart of the Fourier space formalism given by Kaiser was introduced by \citet{Hamilton1992}. Extensions of the linear model, called the ``dispersion model'' have been used to determine the growth rate from the two point galaxy correlation function $\xi(s_{||},s_{\perp})$ \citep[][]{Peacock2001, Hawkins2003}. However, measurements of the growth rate parameter from the dispersion model have been found to introduce systematic errors in the results \citep[][]{Taruya2010, Bianchi2012}. An important breakthrough in the dispersion model was effected by Taruya in 2010 \citep[][]{Taruya2010} when he proposed a new model of redshift space distortion which studied correction factors arising from the non-linear coupling between velocity and density fields. RSD analyses from the data releases 9 \citep[DR09;][]{Ahn2012} and 10 \citep[DR10;][]{Ahn2014} of Sloan Digital Sky Survey III \citep[SDSS III;][]{Eisenstein2011} which include the Baryon Oscillation Spectroscopic Survey \citep[BOSS;][]{Dawson2013}, employed the Lagrangian Perturbation Theory discussed by \citet{Matsubara2008a, Matsubara2008b} and the Gaussian Streaming Model to measure the linear growth rate of structure in the Universe \citep[][]{Reid2011, Reid2012, Samushia2013}. Other measurements of the linear growth rate of the Universe ($f\sigma_8$) include \citet{Cooray2004, Percival2004, Narikawa2010, Blake2011, Giovannini2011, Okumura2011, Beutler2012, Gupta2012, Hirano2012, Hudson2012, Nusser2012, Samushia2012, Shi2012, Contreras2013, delaTorre2013, Macaulay2013, Sanchez2013, Sanchez2014, Reid2014, Avsajanishvili2014, Alam2015a, Feix2015, Marulli2015, Hamaus2016}. 

In the work presented here, we follow \citet{Alam2015a} to study galaxies in the final data release \citep[DR12;][]{Alam2015b}. We use the MultiDark Patchy mock catalogs \citep[][]{Kitaura2014, Kitaura2015} and the BOSS DR12 galaxy dataset \citep[][]{Reid2015} in our analysis to recover information about the growth rate of the Universe at different redshifts. Our paper is a support paper and the final cosmological analysis is discussed in \citet{Acacia2016}. In addition to our paper there are other companion papers of \citet{Acacia2016} which analyze the full shape of the anisotropic two-point correlation function. \citet{Acacia2016} use the methodology presented in \citet{Sanchez2016b} to combine the results of all these companion papers into a final set of BOSS consensus constraints and explore their cosmological implications. In section~\ref{sec:BOSSDR12}, we provide a brief summary of three different full shape analyses of galaxy clustering for the BOSS DR12 sample using different models \citep[][]{Beutler2016a, Grieb2016, Sanchez2016a}, which are support papers to \citet{Acacia2016}. Other companion papers where the BAO scale is measured using the anisotropic two-point correlation function include \citet{Beutler2016b, Ross2016, Vargas2016}.

Our paper is organized as follows. In section~\ref{sec:CLPT}, we review the Convolution Langrangian Perturbation Theory and the Gaussian Streaming Model which we use as the theoretical basis of our investigation. In section~\ref{sec:Data}, we sketch the details of the BOSS DR12 galaxy dataset and the mock galaxy catalogs that we use in our analysis. We discuss details of the approach adopted in our analysis in section~\ref{sec:Analysis}. Our results from the mocks and the galaxy data are discussed in section~\ref{sec:Results}. We conduct a critical analysis and present a summary of the obtained results for cosmological parameters in sections~\ref{sec:Discussion} and~\ref{sec:Summary}.

\section{The CLPT-GSRSD model} \label{sec:CLPT}

\subsection{The Convolution Lagrangian Perturbation Theory}
We choose the theoretical framework for the modeling of correlation functions proposed in the Convolution Lagrangian Perturbation Theory (CLPT) as the theoretical basis of our analysis. The framework of CLPT was proposed by \citet{Carlson2013}. CLPT seeks to use a non-perturbative resummation of the Lagrangian Perturbation Theory to make predictions of correlation functions in real and redshift-space. In CLPT, terms in the expansion of the two point correlation function $\langle \delta_X( \mathbf{x_1} ) \delta_X( \mathbf{x_2} ) \rangle$ for a tracer $X$, which become constants in the limit of large scale, are identified and kept from being expanded in the resummation. We discuss the concept of two point galaxy correlation functions in section~\ref{sec:2PCorr}. The lowest order in the expansion of the matter correlation function, $\xi(s)$ returns the Zel'dovich approximation. We introduce details of the concept of multipoles, i.e. the monopole($\xi_0(s)$) and the quadrupole ($\xi_2(s)$) in equation~\ref{eqn:MultipoleEqn}. The monopole and the quadrupole from CLPT agree well with the results of $N$-body simulations till small scales ($20$ $h^{-1}$Mpc). Also, the results from CLPT tally with the results from the Lagrangian resummation theory (LRT) by Matsubarra \citep[][]{Matsubara2008a, Matsubara2008b} on large scales. Although CLPT has been shown to perform better than LRT and the linear theory in the modeling of multipoles, the desire to model even smaller scales effectively necessitates better performance from the theory, especially with regard to the performance of the quadrupole on smaller scales. 
      
\subsection{The Gaussian streaming model} \label{sec:GSM}
To overcome the deficiencies in the predictions of the linear theory one needs to consider the non-linear mapping between real space and redshift space and the non-linearty of the halo pairwise velocities. The Gaussian streaming model (GSM) attempts to non-perturbatively model the non-linear mapping between real space and redshift space positions in halo redshift-space correlation functions on quasilinear scales ($ \sim 30 - 80$ $h^{-1}$Mpc) \citep[][]{Reid2011}. In the Gaussian streaming model, the pairwise velocity is assumed to have a Gaussian probability distribution function. The mean and the dispersion of the pairwise velocities depend on the pair separation vector and the angle of the pair separation vector with the line of sight. \citet{Reid2011} presented the first calculation of next-to-leading-order corrections to pairwise mean infall velocities and dispersions for linearly biased haloes in their work. The scale dependent Gaussian streaming ansatz predicts the monopole and quadrupole estimators to accuracies of  $\sim 0.5 \%$ above $10$ $h^{-1}$Mpc and $\sim 2\%$ above $25$ $h^{-1}$Mpc respectively for halo correlation functions. However, the Fingers of God effect is expected to affect the observations at scales below $25$ $h^{-1}$Mpc. The multipoles $\xi_{2,4}$ obtained from GSM are significantly more enhanced on quasi-linear scales when compared to the multipoles obtained from the linear theory. Among the more detailed simulation based models which describe the velocity distribution of galaxies are \citet{Zu2013} which analyzes the velocity distribution of galaxies around galaxy groups and \citet{Bianchi2015} which develops a prescription for galaxy pairwise velocities at large scales.

The level of accuracy of the results obtained from GSM leads to it being a desirable candidate for the analysis of two point correlation functions. In order to further improve the estimation of the multipoles from CLPT, the velocity statistics and the correlation function from CLPT are combined with the formalism of GSM to enhance the accuracy of the monopoles and quadrupoles predicted by the theory \citep[][]{Wang2014}.

There are examples of papers in literature \citep[][]{Hawkins2003} where it was found that pairwise velocities have non zero higher order moments and hence, they deviate from perfectly Gaussian distributions. Recently, people have conducted even more detailed studies of pairwise velocity moments, e.g., \citet{Bianchi2015, Bianchi2016} (in simulations) and \citet{Uhlemann2015} (in theory and simulations) and have found that the ensuing profiles of pairwise velocity distributions are not close to Gaussian distributions. Nevertheless, all the aforementioned papers have found that the effect of the non-Gaussian features of velocity moments in correlation functions are pronounced at non-linear scales and negligible at linear scales. Including higher order velocity moments will probably improve the modeling at smaller scales and lead to better signal to noise ratios in RSD measurements. But the evaluation of these components will involve higher order  integrals (which will need to be tested using methods similar to those proposed in \citet{Uhlemann2015} ). We consider such sophisticated tests and modeling beyond the scope of the current paper, and hence, we use just a Gaussian streaming model. This is one of the reasons why we fail to use scales smaller than $25 \ h^{-1}$Mpc in our analysis.

\subsection{CLPT-GSRSD} \label{sec:CLPT-GSRSD}
The idea to use predictions from CLPT along side GSM is motivated by the need to model multipoles more effectively on smaller scales. The use of CLPT to predict the components of GSM was suggested by \citet{Carlson2013} and was implemented by \citet{Wang2014}. 

\begin{align} 
\label{eqn:TheoryCorrFunc}
1+\xi^{\rm model}(s_{||},s_{\perp}) & = \int  \mathrm{exp} \left\lbrace \frac{-\left[ s_{||} - y - \mu v_{12}(r) \right]^2}{2 \left[ \sigma_{12}^2 (r,\mu) + \sigma_{\rm FOG}^2 \right] } \right\rbrace \nonumber \\
& \times \frac{ \left[ 1+\xi(r) \right] }{\sqrt{2 \pi \sigma_{12}^2(r,\mu)}} \ dy
\end{align}

In equation~\ref{eqn:TheoryCorrFunc}, $y$ and $s_{||}$ represent the line of sight separation in real and redshift space respectively. The symbol $s_{\perp}$ denotes the perpendicular separations in both redshift and real space. The parameter $r=\sqrt{y^2+s_{\perp}^2}$ is indicative of the pair separation in real space, while $ \mu = y/r$ depicts the cosine of the angle between the line of sight separation in real space $y$ and the pair separation vector $r$. \citet{Reid2012} presented an extension of the GSM where the parameter $\sigma_{\rm FOG}$ was introduced and tested for the first time as a single extra parameter which deals with the incorporation of the Fingers of God effect in the GSM. The paramter $\sigma_{\rm FOG}$ serves to describe an isotropic dispersion that modifies the scale dependence of quadrupole moments on small distance scales. The parameters $v_{12}(\bm{r})$ and $\sigma_{12}(\bm{r})$ represent the mean infall velocity between pairs of matter tracers and the velocity dispersion along the line of sight respectively. Brief details of the calculation of the parameters $v_{12}(\bm{r})$ and $\sigma_{12}(\bm{r})$ are as follows:

\begin{align}
\label{eqn:TheoryParameters}
v_{12,n}(\bm{r}) &= \left[ 1+\xi(r) \right]^{-1} \int M_{1,n}(\bm{r},\bm{q}) d^3 q  \nonumber \\
\sigma_{12,nm}^2(\bm{r}) &= \left[ 1+\xi(r) \right]^{-1} \int M_{2,nm}(\bm{r},\bm{q}) d^3 q
\end{align}

Here, $M_{1,n}(\bm{r},\bm{q})$ and $M_{2,nm}(\bm{r},\bm{q})$ are specific integrals in CLPT that draw on the linear matter power spectrum $P_{\texttt{lin}}(\mathbf{k})$. The indices $n$ and $m$ are representative of directions along the galaxy position vectors. Details of the form of the integrals $M_{1,n}(\bm{r},\bm{q})$ and $M_{2,nm}(\bm{r},\bm{q})$ are expounded in \citet{Wang2014}. The projection of the mean infall velocity between pairs of matter tracers, $v_{12,n}(\bm{r})$ along the direction of pair separation vector, $\bm{r}$ gives the radial component of $v_{12}$ which is obtained as $v_{12} = v_{12,n} \hat{r}_n$. 

From the projection of the pairwise velocity dispersion in directions along and perpendicular to the pairwise separation unit vector $\hat{r}$, one obtains the parameters $\sigma_{||}^2$ and $\sigma_{\perp}^2$ respectively. Equation~\ref{eqn:Project_Pairwise_vel} shows the scheme in which $\sigma_{12,nm}^2$ is contracted to obtain $\sigma_{||}^2$ and $\sigma_{\perp}^2$. 

\begin{align}
\label{eqn:Project_Pairwise_vel}
\sigma_{||}^2 &= \sigma_{12,nm}^2 \hat{r}_n \hat{r}_m \nonumber \\
\sigma_{\perp}^2 &= \left( \sigma_{12,nm}^2 \delta^K_{nm} - \sigma_{||}^2 \right)/2
\end{align}

Using the knowledge of the components of the pairwise velocity dispersion (i.e. $\sigma_{||}^2$ and $\sigma_{\perp}^2$), one can obtain $\sigma_{12}^2$ as functions of the components $\sigma_{||}^2$ and $\sigma_{\perp}^2$.

\begin{align}
\label{eqn:Pairwise_vel_fin}
\sigma_{12}^2( r, \mu ) = \mu^2 \sigma_{||}^2(r) + \left( 1 - \mu^2 \right) \sigma_{\perp}^2(r)
\end{align}

The combination of CLPT and GSM is shown to improve the accuracy for $\xi_0(s)$ and $\xi_2(s)$ to 2 and 4 percent respectively till scales of $s>25$ $h^{-1}$Mpc. The ensuing model is referred to as the Convolution Lagrangian Perturbation Theory with Gaussian Streaming Redshift Space Distortions (CLPT-GSRSD). We will use the CLPT-GSRSD model to derive deductions about constraints on cosmological parameters including the growth rate. 

\section{Data} \label{sec:Data}
\begin{figure*}
    \includegraphics[width=0.32\textwidth]{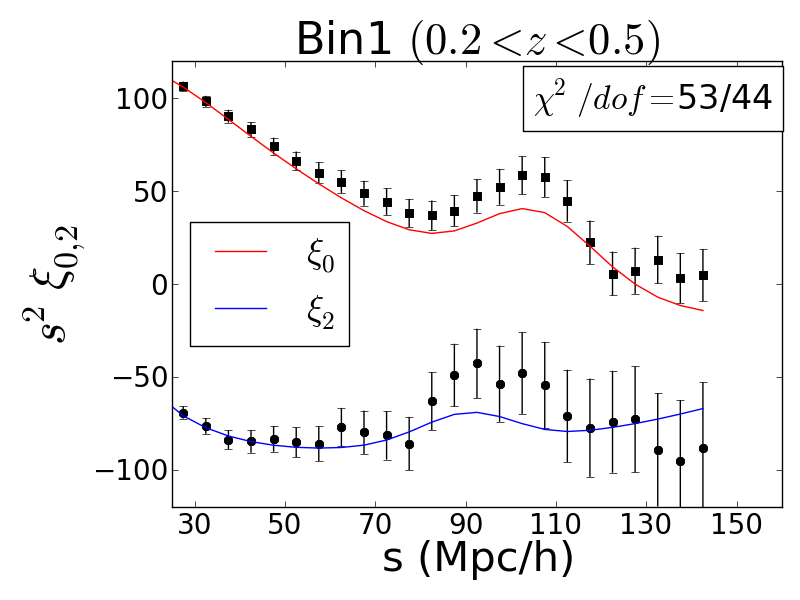}
    \includegraphics[width=0.32\textwidth]{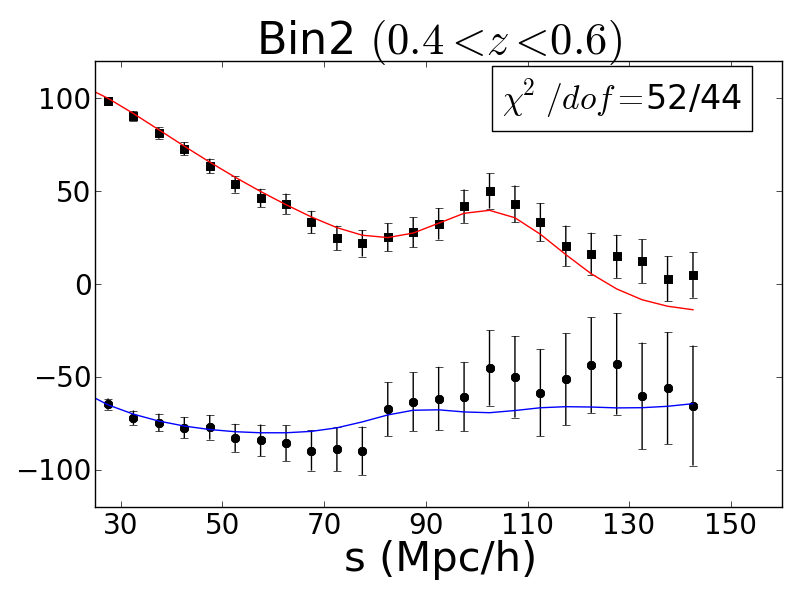}
    \includegraphics[width=0.32\textwidth]{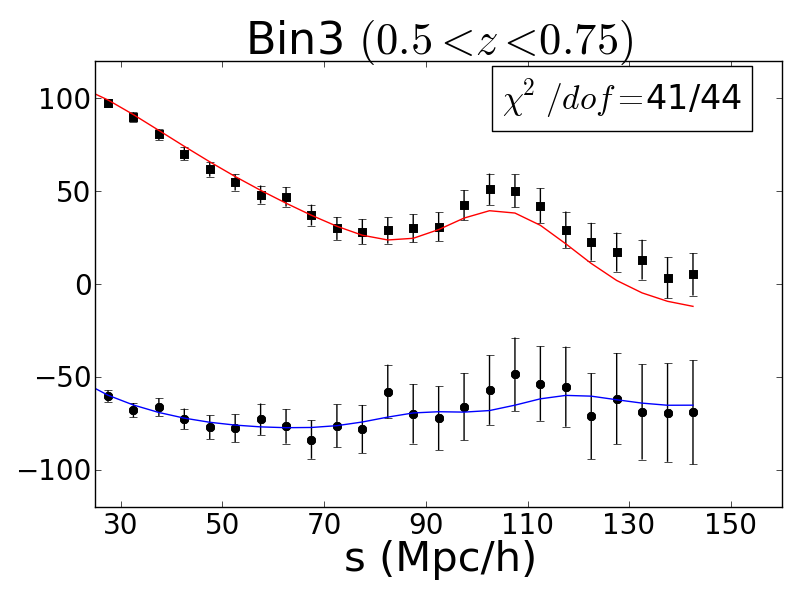}        
    \caption{The black dots in the plots of this figure represent monopole ($_{^{_{\blacksquare}}}$) and quadrupole ($\bullet$) for BOSS DR12 galaxy sample evaluated at different values of $s$. The error bars are obtained from the diagonal elements of the covariance matrices corresponding to mocks in the three redshift bins. The red and the blue lines denote the best fit models of monopole and quadrupole of the galaxy data. The analysis assumes a fitting range $25 \ h^{-1}$Mpc$ \leq s \leq 150 \ h^{-1}$Mpc with a bin size of $5 \ h^{-1}$Mpc.}
    \label{fig:CombBin}
\end{figure*}

In this section we sketch the nuances of the galaxy and the mock datasets that we use in our analysis. We briefly touch upon the particulars of the BOSS DR12 galaxy dataset and give necessary details of the Multi-Dark Patchy mock catalogs.

\subsection{The BOSS DR12 Galaxy Dataset} \label{sec:BOSSDR12}
We use the Data Release 12 \citep[DR12;][]{Alam2015b} of the Sloan Digital Sky Survey III \citep[][]{Eisenstein2011} Baryon Oscillation Spectroscopic Survey \citep[][]{Dawson2013} dataset in our analysis. The dataset includes an ensemble of galaxies obtained from the SDSS survey using a wide-field, drift-scanning CCD camera \citep[][]{Gunn1998, Gunn2006} and selected using multicolor imaging photometry in five color bands \citep[$u,g,r,i,z$;][]{Fukugita1996}). This dataset was used by BOSS multi-fibre spectographs \citep[][]{Bolton2012, Smee2013} to observe the spectra of 1,198,006 galaxies. These galaxies were observed along with approximately 300,000 quasars, 200,000 stars and 400,000 ancillary objects. The observations were made in a sequence of 15-minute exposures and integrated until a minimum signal-to-noise ratio is reached for the faint galaxy targets. This procedure ascertains homogeneity in the dataset with a redshift completeness of over 97\% for the entire survey footprint. \citet{Bolton2012} describe the method for the determination of redshift from the classified spectra.

The selected sample of 1,198,006 galaxies encompasses a redshift range of $0.2$ to $0.75$ and covers 9,329 square degrees. For the purpose of this paper, we divide this redshift range into three overlapping redshift bins of roughly equal volume \citep[][]{Acacia2016}, $viz.$  $0.2<z<0.5$ (\textit{bin1}), $0.4<z<0.6$ (\textit{bin2}) and $0.5<z<0.75$ (\textit{bin3}). These bins have effective redshifts of $z_{\rm eff} = 0.38,0.51$ and $0.61$ respectively. For the three bins, we work with a $\Lambda$CDM-GR cosmological model with a fiducial cosmology of $\Omega_m = 0.31, \ H_0 = 0.676, \ \Omega_{\Lambda} = 0.69, \ \Omega_b h^2 = 0.022$ and $\sigma_8 = 0.80$. 

We follow the methods outlined in \citet{Ross2012} and \citet{Anderson2014} to give weights to each galaxy under study to compensate for the effects of redshift failures and fiber collisions. We introduce the weight factor $w_{\rm zf}$ to account for redshift failure of the nearest neighbor of a galaxy. Similarly, the weight factor $w_{\rm cp}$ is intended to account for a scenario where the redshift of a neighbor was not obtained because it was in a close pair. The weight factor $w_{\rm star}$ serves to correct for the non-cosmological fluctuations which arise due to the dependence of target identification on the local star density. The weight factor $w_{\rm see}$ corrects the effect of the seeing conditions (during photometric observations) on the target density. The factor $w_{\rm FKP}$ is included because of the need to minimize the variance in the weighted number of galaxy counts. Following is the weighting scheme that we use: $ w_{\rm tot} = \left( w_{\rm cp} + w_{\rm zf} - 1 \right) w_{\rm star}w_{\rm see}w_{\rm FKP} $.

Four companion papers (including this paper) present different approaches for the full-shape analysis of the BOSS DR12 combined galaxy sample \citep[][]{Acacia2016}:
\begin{enumerate}
  \item In this work, we use multipoles obtained from anisotropic two-point galaxy correlation functions to analyze the DR12 data. Details of the approach used for our analysis are given in section~\ref{sec:Analysis}. 
  \item \citet{Sanchez2016a} present an analysis of the BOSS DR12 combined galaxy sample using wedges obtained from anisotropic two-point correlation functions. 
  \item The methodology presented in \citet{Beutler2016a} for the analysis of DR12 galaxy data employs multipoles obtained from anisotropic power spectrum.
  \item \citet{Grieb2016} use an analysis based on wedges from anisotropic power spectrum in their investigation of the BOSS DR12 galaxy data.
\end{enumerate}

\subsection{Mock Galaxy Catalogs}
We use the Multi-Dark Patchy (MD-P) mock catalogs \citep[][]{Kitaura2014, Kitaura2015} as an essential statistical tool and as a precursor to the analysis of the SDSS III DR12 combined galaxy dataset. These mock catalogs require the generation of accurate reference catalogs. For these MD-P mock catalogs, the reference catalogs are extracted from one of the BigMultiDark cosmological N-body simulations \citep[][]{Klypin2014} which uses gadget-2 \citep[][]{Springel2005} with 3840$^3$ particles in a volume of $( 2.5 \ h^{-1}$Mpc$)^3$. These simulations are based on a $\Lambda$CDM cosmology of $H_0=67.77 \ \texttt{km.s}^{-1}.\texttt{Mpc}^{-1}, \ \Omega_m=0.307115, \ \Omega_b=0.048206, \ n_s=0.9611$ and $\sigma_8=0.8288$.  

In a manner akin to the division of the BOSS DR12 data into redshift bins, the MD-P mock catalogs that we use are segregated into three redshift bins with effective redshifts of $z_{\rm eff} = 0.38$ (\textit{bin1}), $0.51$ (\textit{bin2}) and $0.61$ (\textit{bin3}). In each redshift bin we use 997 mocks in our analysis. The primary purpose of the use of the MD-P mocks is to assist in the formulation of covariance matrices for the different bins of the galaxy dataset and to mimic the statistics of the same. We discuss more about the use of the MD-P mocks to obtain covariance matrices in sections~\ref{sec:2PCorr} and~\ref{sec:Covariance}.

\section{Analysis} \label{sec:Analysis}
In this section, we outline the methodology that we have used in our analysis of the MultiDark-Patchy (MD-P) mock catalogs and  the BOSS DR12 dataset in the three redshift bins. We sketch the steps that we employ in analyzing the positions of galaxies to obtain multipoles ($\xi_0(r)$ and $\xi_2(r)$) from two-point correlation functions ($\xi(r)$). We also discuss the computation of covariance matrices from MD-P mock catalogs. We conclude by shedding light on the use of Markov Chain Monte Carlo (MCMC) in chosen parameter spaces for the MD-P mocks and the SDSS III galaxy dataset to obtain a handle on the variation of different RSD and BAO parameters.  

\subsection{The two-point galaxy correlation function} \label{sec:2PCorr}
In the fiducial cosmology $\Omega_m = 0.31, \ H_0 = 0.676, \ \Omega_{\Lambda} = 0.69, \ \Omega_b h^2 = 0.022$ and $\sigma_8 = 0.80$, we map redshift and celestial coordinates ($\alpha, \delta$) to the position of a galaxy in three dimensional space. We use the Landy-Szalay estimator \citep[][]{Landy1993} to obtain the two-point correlation function $\hat{\xi}(s)$ for a given galaxy sample.  

\begin{equation}
\hat{\xi}_{\rm LS}(s, \mu) = \frac{DD(s, \mu) - 2DR(s, \mu) + RR(s, \mu)}{RR(s, \mu)}
\end{equation}

Here $\mu=cos \ \theta$ (where $\theta$ is the angle between the line of sight and the radial distances), $DD(s, \mu)$ is the pair count of galaxies with separation $s$ and orientation $\mu$, $DR(s, \mu)$ is the cross-pair counts between the galaxies and a random distribution, and $RR(s, \mu)$ is the number of pairs for a random distribution. The Landy-Szalay estimator has only a second order bias caused by finite sample effects. The performance of the Landy-Szalay estimator has been proved to be better than other comparable two point correlation functions at large scales \citep[][]{Pons1999, Kerscher2000}. In the measurement of the two-point correlation function, each galaxy pair is weighted by $w_{\rm tot}$. More details of the weighting scheme that we use can be found in section~\ref{sec:BOSSDR12}. 

The use of two dimensional two-point correlation functions ($ \hat{\xi} (s, \mu )$) will lead to a large number of bins (due to the presence of two dimensions). To fit a two-dimensional two-point correlation function directly, we will need to construct a large covariance matrix which will in turn necessitate the use of a very large number of mocks. Creating such a large number of mocks will be computationally intensive. This motivates a need to reduce the number of bins. We use different kernels to obtain averages of the two dimensional two-point correlation function ($\hat{\xi}(s, \mu)$) to condense the information contained in it.

\begin{equation}
\label{eqn:Isotropized2PCorr}
\tilde{\xi}_{f}(s) = \int f(s, \mu) \hat{\xi}( s, \mu ) dV
\end{equation} 

Here $f(s, \mu)$ is an appropriately selected kernel function. We outline two different ways of defining these kernels ; one, where we take recourse to the use of clustering wedges, $\tilde{\xi}_{\Delta \mu}(s)$, and another, where we use angle averaged correlation functions, $\tilde{\xi}(s)$. 

In the first approach, one obtains a clustering wedge by averaging the two dimensional two-point correlation function ($ \hat{\xi} (s, \mu )$) over a chosen interval $\Delta \mu = \mu_{\rm max} - \mu_{\rm min}$ \citep[][]{Kazin2012, Sanchez2013, Sanchez2014}, i.e. 

\begin{align}
\label{eqn:Wedge}
\tilde{\xi}_{\Delta \mu}(s) &\equiv \frac{1}{ \left(  \mu_{\rm max} - \mu_{\rm min} \right) } \int_{\mu_{\rm min}}^{\mu_{\rm max}}  \hat{\xi}_{\rm LS}( s, \mu ) d \mu \nonumber \\
&\approx  \frac{1}{ \left(  \mu_{\rm max} - \mu_{\rm min} \right) } \displaystyle\sum_{j} \Delta \mu_j \hat{\xi}_{\rm LS}(s,\mu_j)
\end{align}

In the second approach, we obtain isotropized correlation functions by finding projections of the anisotropic two dimensional two-point correlation functions in the basis of Legendre polynomials. \citet{Hamilton1992} demonstrated one of the first uses of spherical harmonics as kernels when he used Legendre polynomials to obtain averages of the anisotropic $\xi(s, \mu)$. In our analysis, we also use the orthonormal basis of Legendre polynomials $P_{\ell}(\mu)$ as our kernel functions to obtain `multipoles' of different orders, $\ell$ ($ \ell = 0,1,2 \ \ldots$). 

\begin{align}
\label{eqn:MultipoleEqn}
\tilde{\xi}_{\ell}(s) &= \frac{2 \ell + 1}{2} \int^{1}_{-1}\hat{\xi}_{\rm LS}(s,\mu) P_{\ell}(\mu)d\mu \nonumber \\
&\approx  \frac{2 \ell + 1}{2} \displaystyle\sum_{j} \Delta \mu_j \hat{\xi}_{\rm LS}(s,\mu_j) P_{\ell}(\mu_j)
\end{align}

Due to the anti-symmetry of Legendre polynomials of odd orders, the angle averaged 2-point correlation function of odd orders ($\tilde{\xi}_{2 \ell + 1}(s)$) vanish. We use even order multipoles i.e., monopole ($\tilde{\xi}_{0}(s)$) and the quadrupole ($\tilde{\xi}_{2}(s)$) for our analysis. One observes huge errors for multipoles of the fourth order, $viz.$ hexadecapoles ($\tilde{\xi_4}(s)$). Hexadecapoles from the theoretical model are suspected to be more susceptible to survey systematics and the correction schemes adopted in the model than $\tilde{\xi_0}(s)$ and $\tilde{\xi_2}(s)$. Also, for hexadecapoles, one sees disagreement between the results of CLPT and $N$-body simulations \citep[][]{Carlson2013}. Hence, we base our analysis on multipoles of orders no higher than 2. 

The datasets that we use have 120 bins in $\mu$. We use equation~\ref{eqn:MultipoleEqn} to compute multipoles corresponding to the MD-P mocks and the BOSS DR12 dataset in evenly spaced bins of width $5$ $h^{-1}$Mpc in $s$. Our analysis has been found to be relatively insensitive to bin sizes for bin sizes between $2-8$  $h^{-1}$Mpc \citep[][]{Alam2015a}. Since these multipoles are computed from data, we will use the notation $\xi^{\rm data}$ to refer to these multipoles. We bin the correlation function multipoles obtained from theory into the same bins as that of data for the purpose of comparison of the multipoles obtained from data with those obtained from theory. In the first step, we calculate $\xi_0(s)$ and $\xi_2(s)$ for the MD-P mocks in each redshift bin using the information about galaxy correlation functions $\xi(s, \mu)$ at different positions. Figure~\ref{fig:CombBin} illustrates the monopole and the quadrupole for BOSS DR12 galaxy data and the best fit models of the data for the three bins. More details about the techniques used to obtain the error bars in Figure~\ref{fig:CombBin} are given in section~\ref{sec:Covariance} while sections~\ref{sec:CosmoMC} and~\ref{sec:TheoryMethods} describe the methods used to obtain the best fit models. In section~\ref{sec:BOSSDR12GalaxyData} we give a comprehensive description of the details of Figure~\ref{fig:CombBin}.   

\begin{figure}
    \centering
    \includegraphics[width=0.45\textwidth]{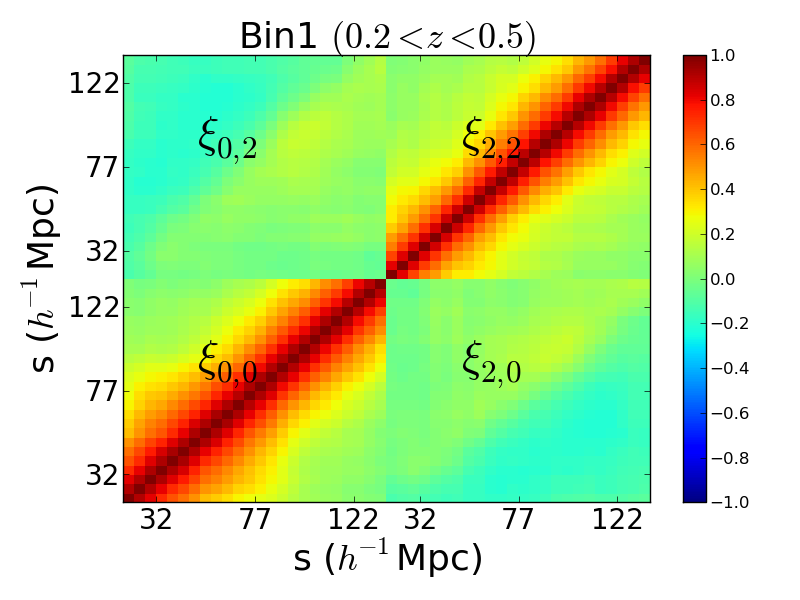}
    \includegraphics[width=0.45\textwidth]{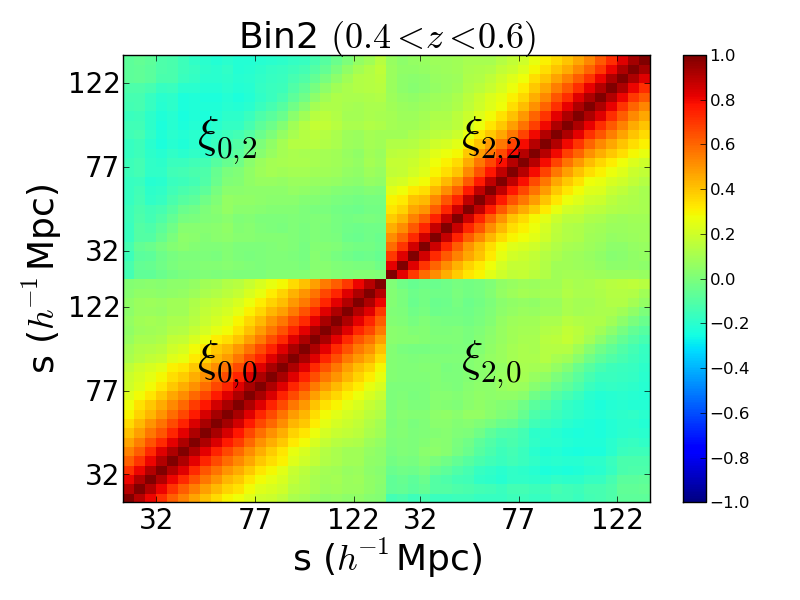}
    \includegraphics[width=0.45\textwidth]{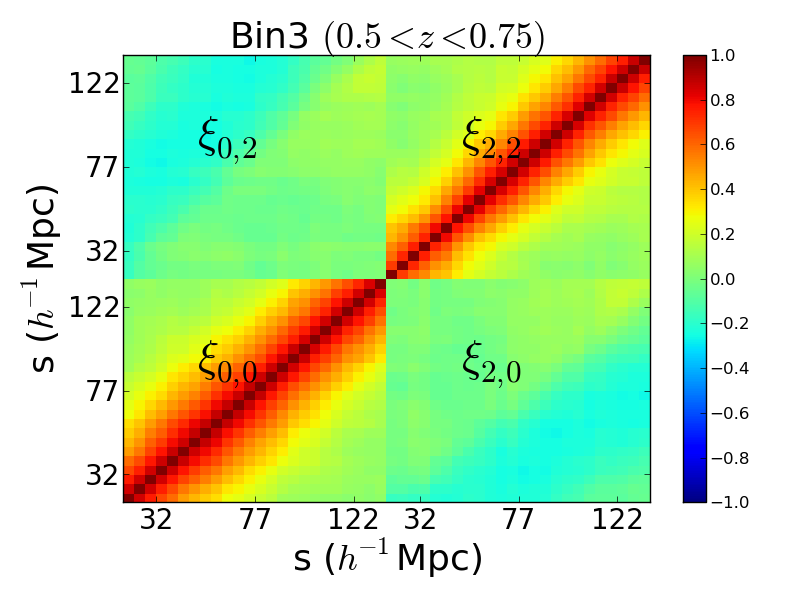}
    \caption{The plot on the top of the figure shows the correlation matrices obtained from MD-P mocks with $z_{\rm eff}=0.38$. The plot in the center denotes the correlation matrix gotten from MD-P mocks with $z_{\rm eff}=0.51$. The plot on the bottom of the figure shows the correlation matrix obtained from MD-P mocks with $z_{\rm eff}=0.61$.}
    \label{fig:CorrMatrices}
\end{figure}

\subsection{The covariance matrix} \label{sec:Covariance}
We use the multipoles corresponding to the MD-P mocks in a given bin to get an estimate of the covariance matrix of that bin. We follow the prescription outlined in \citet{Vargas2013, Percival2013} to calculate the covariance matrix from the monopole and the quadrupole corresponding to the mocks.  

\begin{equation}
\label{eqn:CovarianceEqn}
\hat{\Sigma}_{ij} = \left[ \displaystyle\sum_{n=1}^{N_{\rm mock}}(\tilde{\xi}_{i,n} - \bar{\xi})(\tilde{\xi}_{j,n} - \bar{\xi}) \right] \Bigg/ \left[ N_{\rm mock}-1 \right]
\end{equation}

In equation~\ref{eqn:CovarianceEqn}, $\hat{\Sigma}_{ij}$ is the $(i,j)$th entry of the calculated covariance matrix, where the indices `$i,j$' correspond to the index of the binned value of the radial position, $s$ in the two point correlation function. $N_{\rm mock}$ is the total number of mocks and the index `$n$' denotes the number (index) of a given mock. $\bar{\xi}$ corresponds to the mean of the mocks. The correlation matrix, $\hat{\texttt{r}}$ is obtained from the covariance matrix $\Sigma$ as

\begin{equation}
\hat{\texttt{r}}_{ij} = \frac{\hat{\Sigma}_{ij}}{ \sqrt{ \hat{\Sigma}_{ii}\hat{\Sigma}_{jj} }}
\end{equation}

The inverse of the covariance matrix obtained from the mocks is used to estimate the likelihood as shown in equation~\ref{eqn:chi2}.

Figure~\ref{fig:CorrMatrices} shows the correlation matrices obtained from the three redshift bins of the MD-P mock catalogs, i.e. $z_{\rm eff}=0.38, \ 0.51$ and $0.61$.

\subsection{The \textsc{CosmoMC} code and parameterizations} \label{sec:CosmoMC}
We compare the results for multipoles that we obtain for data (or mock catalogs) using equation~\ref{eqn:MultipoleEqn} with the correlation functions that we obtain from theory using equation~\ref{eqn:TheoryCorrFunc} in a Markov Chain Monte Carlo (MCMC) analysis which operates in a chosen parameter space.  We use \textsc{CosmoMC} \citep[][]{Lewis2002} to effectuate the MCMC algorithm to explore the chosen cosmological parameter space. For the MD-P mocks we use a six dimensional parameter space with $\lbrace F_1, F_2, f, \sigma_{\rm FOG}, \alpha, \epsilon \rbrace$ being the parameters in the MCMC analysis. For the three bins in the BOSS DR12 galaxy data, we explore a ten dimensional parameter space: $\lbrace \Omega_b h^2, \ \Omega_c h^2, \ n_s, \ \ln (10^{10} A_s), \ F_1, \ F_2, \ f, \ \sigma_{\rm FOG}, \ \alpha, \ \epsilon \rbrace$ in the MCMC analysis. Here $\Omega_b, \ \Omega_c$ represent the baryon density and the dark matter density respectively and the symbol $h$ denotes $H_0/100$ where $H_0$ is the Hubble constant. The parameter $A_s$ is the amplitude of the primordial power spectrum. The quantities $F_1, \ F_2$ are the first and the second order Lagrangian biases while the quantity $f$ depicts the logarithm derivative of the growth factor, i.e. $f = d \ln (D) / d \ln (a)$. For $\Lambda$CDM cosmology, the growth factor $f$ can be related to the matter density $\Omega_m$ by the approximation $f(a) \approx \Omega^{0.55}(a)$. It is common to report measurements of the linear growth rate in terms of $f \sigma_8$. Here $\sigma_8$ corresponds to the rms fluctuation of mass within a top-hat sphere of $8 \ h^{-1}$Mpc  radius and it is treated as a normalization constant. The symbol $\sigma_{\rm FOG}$ accounts for an isotropic velocity dispersion due to the Fingers of God effect.     

The comoving galaxy power spectrum contains cosmological information about galaxy clustering. In practice, we measure galaxy redshifts and angles and deduce cosmological distances from these instead of measuring clustering directly in comoving space. This approach relies on the use of the relevant cosmological model to convert redshift to cosmological distance. The use of erroneous or inaccurate cosmological models leads to distortions in the comoving clustering and incorrect measurement of distances. These cosmological distortions are manifested in the radial and the angular directions of the clustering signal. Such distortions are sensitive to the Hubble parameter ($1/H(z)$) and the angular diameter distance ($D_{\rm A}(z)$) in the radial and the angular directions respectively. The Alcock-Paczy\'{n}ski (AP) test  \citep[][]{Alcock1979, Lopez2014} presents an approach where one tries to mitigate these distortions by adjusting the cosmological model. The primary advantage of the AP approach and the use of AP parameters is that they depend only on the geometry of the Universe. The parameter $\alpha$ is representative of the assessment of BAO from spherically averaged clustering measurements while the parameter $\epsilon$ illustrates the importance of the BAO feature in $\xi_2$ for a spherically symmetric sample. When a fiducial cosmological model is used to infer distances from the redshift, the separation between the observed and the expected positions of the BAO pertains to two dilation scales. The dilation scale which is parallel to the line of sight is denoted by $\alpha_{||}$ while the dilation scale perpendicular to the line of sight is represented by $\alpha_{\perp}$. The parameters $\alpha_{||}$ and $\alpha_{\perp}$ can be obtained from knowledge of the Hubble expansion rate, $H(z)$ and the angular diameter distance, $D_{\rm A}(z)$ for the fiducial and model cosmologies.

\begin{equation}
\label{eqn:Alcock_Parameters_1}
\alpha_{||} = \frac{H^{\rm fid}(z)r^{\rm fid}_{\rm s}(z_{\rm d})}{H(z)r_{\rm s}(z_{\rm d})}, \ \ \ \ \alpha_{\perp} = \frac{D_{\rm A}(z)r^{\rm fid}_{\rm s}(z_{\rm d})}{D^{\rm fid}_{\rm A}(z)r_{\rm s}(z_{\rm d})}
\end{equation}

Here, the superscript `fid' refers to the value of a quantity in the fiducial cosmology and the parameter $r_{\rm s}(z_{\rm d})$ denotes the fiducial sound horizon assumed in the power spectrum template. The parameter $r_{\rm s}(z_{\rm d})$ sets the comoving BAO scale. It is also common to obtain the $\alpha_{||}$ and $\alpha_{\perp}$ as derived parameters from $\alpha$ and $\epsilon$ using the following relationships:

\begin{equation}
\label{eqn:Alcock_Parameters_2}
\alpha = \alpha_{||}^{1/3} \alpha_{\perp}^{2/3}, \ \ \ \ 1 + \epsilon = \left( \frac{\alpha_{||}}{\alpha_{\perp}} \right)^{1/3}
\end{equation} 

One can also use the volume averaged distance $D_{\rm v}$ and the AP-parameter $F_{\rm AP}$ to report measurements of the angular and radial projected distance scales. The volume averaged distance $D_{\rm v}$ is related to the redshift $z$, the speed of light $c$, the angular diameter distance $D_{\rm A}$ and the Hubble constant $H$ by the following relation:

\begin{equation}
\label{eqn:D_V}
D_{\rm v} = \left[ (1+z)^2 c z \frac{D_{\rm A}^2}{H} \right]^{1/3}
\end{equation} 

The AP-parameter $F_{\rm AP}$ is defined as:

\begin{equation}
\label{eqn:F_AP}
F_{\rm AP} = \frac{1+z}{c} D_{\rm A} H
\end{equation}

In the work presented in this paper, we allow the AP effect to change due to cosmology through the free parameters $\alpha$ and $\epsilon$. This essentially accounts for error in cosmology, which is small, and hence, slightly inflates our distance measurement error bars. But, this is largely ignored in other analyses where the cosmology is kept fixed while allowing the AP parameters to be free. Also, we allow the shape of the power spectrum to vary using $A_s$ and $n_s$ which gives $\sigma_8$ indirectly. This is in contrast with the methods used in other analyses which fix $A_s$ and $n_s$ and vary $\sigma_8$ directly at lower redshifts. This could affect the measurement of growth rate ($f$) and $\sigma_8$ but $f\sigma_8$ is independent of this choice in the parameter space.
 
Once the choice of parameters is fixed, \textsc{CosmoMC} executes a MCMC algorithm to sample the parameter space and compute the linear power spectrum $P_{\texttt{lin}}(\mathbf{k})$ for the sampled points in the parameter space using the Code for Anisotropies in the Microwave Background (\textsc{Camb}) \citep[][]{Lewis2000}. \textsc{Camb} takes the cosmological parameters of the model that we are working with as input. At each sampled point, CLPT uses the linear power spectrum to compute the velocity statistics and the two-point correlation function, $\xi(r)$. From this two-point correlation function, the Gaussian streaming model GSRSD calculates the redshift space two-point correlation function. This is followed by the rescaling of the two-point correlation function in relation to the difference in the fiducial and the current MCMC cosmologies to obtain the correlation function corresponding to the theory, $\xi^{\rm model}$. We follow the approach adopted in \citet{Marulli2012, Xu2012, Samushia2014, Alam2015a} to rescale the model redshift space correlation function to account for the cosmological distortions in the clustering signal.

\begin{equation}
\label{eqn:Rescaling}
\xi^{RSD}(s_{\parallel},s_{\perp}) = \xi^{\rm model} ( \alpha_{\parallel} s_{\parallel}, \alpha_{\perp} s_{\perp} ) 
\end{equation}

Here, $s_{\parallel}$ represents the separation between two galaxies along the line of sight while $s_{\perp}$ represents the separation between two galaxies perpendicular to the line of sight. The rescaling discussed in equation~\ref{eqn:Rescaling} is a result of the Alcock-Paczy\'{n}ski effect and it involves the use of the Alcock-Paczy\'{n}ski parameters which are discussed in equation~\ref{eqn:Alcock_Parameters_1}. Corresponding to each set of cosmological parameters in the sampling process we get theoretical models for monopole and quadrupole using the aforementioned prescription. More details of the evaluation of multipoles from two point correlation functions using CLPT are given in section~\ref{sec:TheoryMethods}. A comparison of the multipoles obtained from the theory and the data yields the likelihood. Optimization of the obtained likelihood helps us arrive at the best fit model corresponding to a given data (section~\ref{sec:Likelihood}). 

\subsection{Correlation functions from CLPT-GSRSD} \label{sec:TheoryMethods}
A comprehensive analysis of the chosen cosmological parameters will necessitate the comparison of the the two point correlation functions obtained from data~\ref{sec:2PCorr} with those obtained from theory. We use CLPT-GSRSD to evaluate the required theoretical two point correlation functions ($\xi^{\rm model}$). We discuss the Convolution Lagrangian Perturbation Theory in section~\ref{sec:CLPT}. Details of the incorporation of the Gaussian streaming model into this formalism is described in~\ref{sec:GSM}. We present a brief sketch of the CLPT-GSRSD model in section~\ref{sec:CLPT-GSRSD}. Equation~\ref{eqn:TheoryCorrFunc} encapsulates the details of the technique used to calculate the theoretical correlation functions. We rely heavily on the work presented in \cite{Wang2014} for this section.     

\subsection{Likelihood analysis} \label{sec:Likelihood}
Concatanated combinations of the linearly independent constructs of monopoles ($\xi_0$) and quadrupoles ($\xi_2$) can be thought of as vectors. We assume the correlation function multipoles to be Gaussian distributed. Also, we ignore the parameter dependence of the covariance matrices in our analysis. Consequently, the check of the extent of correspondence between the data and the model vectors and the determination of the likelihood of the parameter values reduces to the computation of the $\chi^2$. Given a data vector $\xi^{\rm data}$, a model vector $\xi^{\rm model}$ and the covariance matrix $\Sigma$, the $\chi^2$ is obtained as

\begin{equation}
\label{eqn:chi2}
\chi^2 = \left( \xi^{\rm data} - \xi^{\rm model} \right) \Sigma^{-1} \left( \xi^{\rm data} - \xi^{\rm model} \right)^T
\end{equation}

In our analysis $\xi^{\rm data} = \left[ \tilde{\xi}^{\rm data}_{0}, \ \tilde{\xi}^{\rm data}_{2} \right]$ and $\xi^{\rm model} = \left[ \tilde{\xi}^{\rm model}_{0}, \ \tilde{\xi}^{\rm model}_{2} \right]$ and $\Sigma^{-1}$ is the inverse of the covariance matrix obtained from the MD-P mocks. 

The RSD likelihood, $\mathcal{L}_{\rm RSD}$ of a parameter $p$ is now given by: $\mathcal{L}_{\rm RSD}(p) \propto e^{-\chi^2(p)/2}$

%
Cosmic Microwave Background (CMB) data from the Planck satellite puts constraints on the cosmological parameters. The Planck temperature anisotropy data from \citet{Planck2014b} is used to compute the Planck likelihood. We use the covariance obtained from \citet{Planck2014c}. To get the full Planck likelihood, we use a multi-variate Gaussian approximation. This Gaussian approximation is close to the likelihood in the parameter space that we are doing our analysis in. The four cosmological parameters mentioned in equation given below, i.e. $\lbrace \Omega_{\rm b}h^2, \ \Omega_{\rm c}h^2, \ n_{\rm s}, \ \ln (10^{10} A_{\rm s}) \rbrace$ represent the likelihood really well and provide an excellent speed gain.

\begin{align*}
\label{eqn:PlanckResults}
& \Omega_{\rm b}h^2 =  0.02207, \ \Omega_{\rm c}h^2 = 0.1196, \ n_{\rm s}=0.9616, \nonumber \\
& \ln (10^{10} A_{\rm s})=3.098, \nonumber \\
& \Sigma_{\rm Planck}  =  \ \ \ \ \ \ \ \ \ \ \ \ \ \ \ \ \ \ \ \ \ \ \ \ \ \ \ \ \ \ \ \ \ \ \ \ \ \ \ \ \ \ \ \ \ \ \ \ \ \ \ \ \ \ \ \ \ \ \ \ \ \ \    \nonumber \\ 
& \begin{tiny} 
\begin{pmatrix} 1.089 \times 10^{-7} & -4.501 \times 10^{-7} & 1.365 \times 10^{-6} & 3.564 \times 10^{-6} \\ -4.501 \times 10^{-7} & 9.610 \times 10^{-6} & -2.215 \times 10^{-5} & 1.562 \times 10^{-5} \\ 1.365 \times 10^{-6} & -2.215 \times 10^{-5} & 8.836 \times 10^{-5} & 2.030 \times 10^{-5} \\ 3.564 \times 10^{-6} & 1.562 \times 10^{-5} & 2.030 \times 10^{-5} & 5.184 \times 10^{-3} \end{pmatrix} 
\end{tiny} 
\end{align*} 

\begin{figure}
    \includegraphics[width=0.41\textwidth]{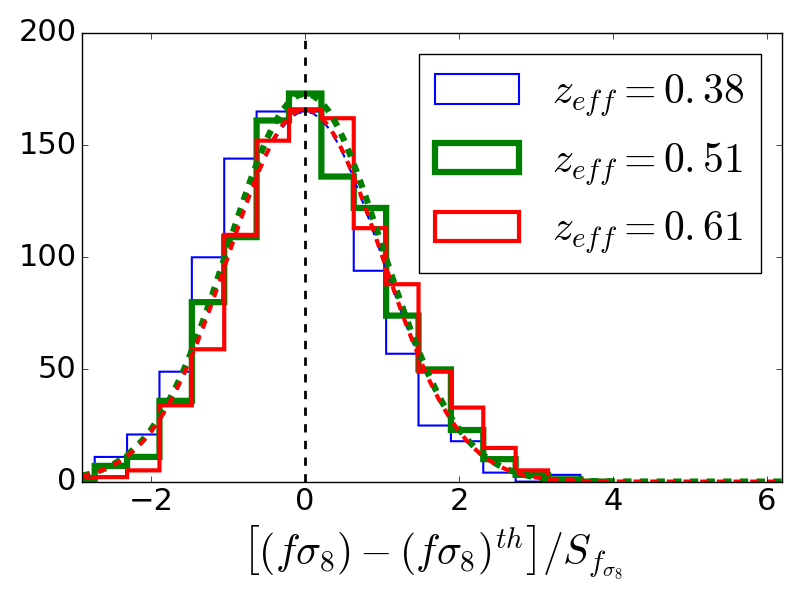}
    \includegraphics[width=0.41\textwidth]{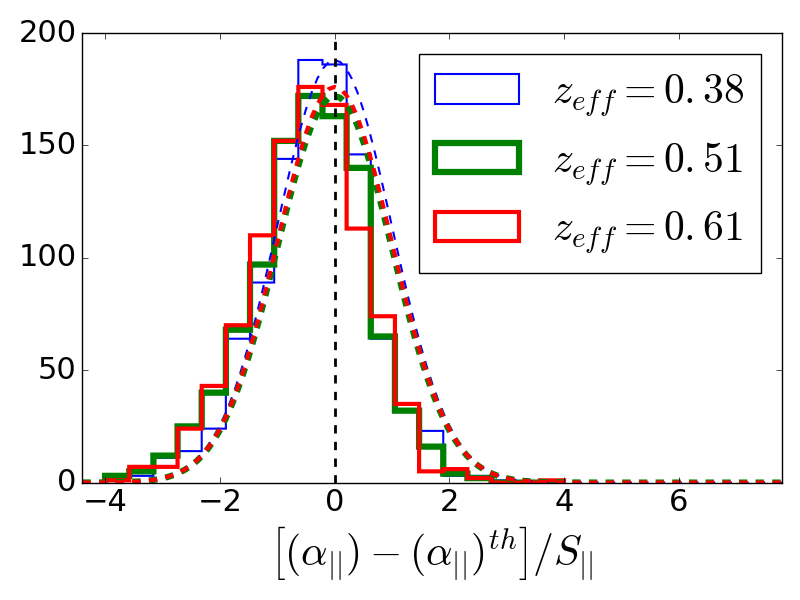}
    \includegraphics[width=0.41\textwidth]{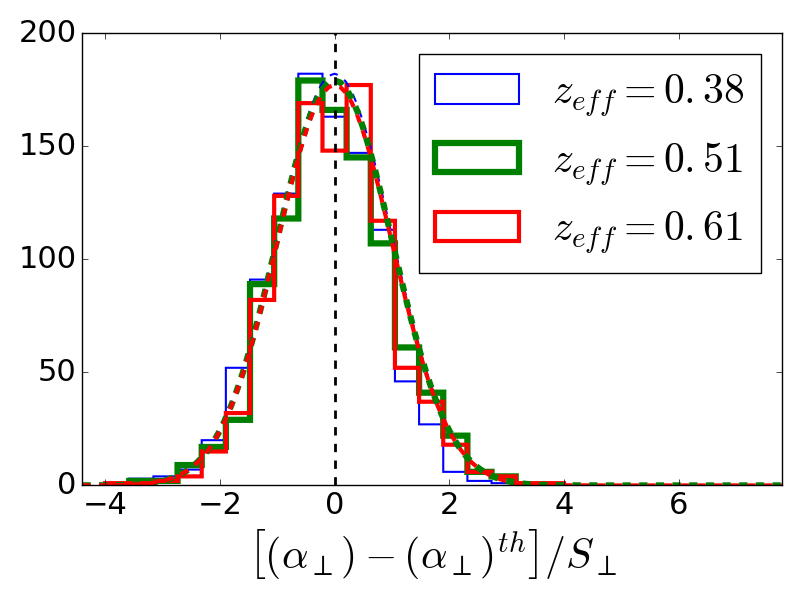}
    \caption{Histograms of the distributions of parameters obtained from the analysis of the MD-P mocks for the three bins. We independently fit the theory to the correlation function multipoles for each of the 997 mocks (in each bin) using the covariance matrices obtained from the mocks to obtain the statistics shown in these figures. The solid blue line represents the bin corresponding to $0.2<z<0.5$ whereas the solid green line depicts the bin for the redshift range $0.4<z<0.6$. The solid red line denotes the bin with the redshift range $0.5<z<0.75$. The dashed blue, green and red lines depict Gaussian functions with zero means, unit variances and heights equal to the heights of the histograms corresponding to $z_{\rm eff} = 0.35, \ 0.51$ and $0.61$ respectively. The symbols $S_{f \sigma_8}$, $S_{||}$ and $S_{\perp}$ represent the standard deviations of the parameters $f\sigma_8$, $\alpha_{||}$ and $\alpha_{\perp}$ respectively. The x-axes of the three plots denote the ratio of the differences between the obtained parameters $f \sigma_8, \ \alpha_{||}, \ \alpha_{\perp}$ and their respective theoretical values $(f \sigma_8)^{th}, \ (\alpha_{||})^{th}, \ (\alpha_{\perp})^{th}$ with their respective standard deviations $S_{f \sigma_8}$, $S_{||}$, $S_{\perp}$. For \textit{bin1}, we have $(f \sigma_8)^{th}=0.484, \ (\alpha_{||})^{th}=1.0031, \ (\alpha_{\perp})^{th}=1.0008$. For \textit{bin2}, we find $(f \sigma_8)^{th}=0.483, \ (\alpha_{||})^{th}=1.0040, \ (\alpha_{\perp})^{th}=1.0010$. For \textit{bin3}, we have $(f \sigma_8)^{th}=0.477, \ (\alpha_{||})^{th}=1.0046, \ (\alpha_{\perp})^{th}=1.0012$.}
    \label{fig:PatchyHistogram}
\end{figure}

We multiply the Planck likelihood with the RSD likelihood to get a joint constraint on the parameters in our analysis: $\mathcal{L}_{\rm total}(p) = \mathcal{L}_{\rm Plack}(p) \times \mathcal{L}_{\rm RSD}(p) $.

For each set of cosmological parameters ($p$) we obtain $\chi^2(p)$. Choice of parameters corresponding to the minimum $\chi^2(p)$ leads to maximization of the likelihood $\mathcal{L}(p)$ and selection of the best fit model. This analysis is performed independently for data in all effective redshift bins. The results obtained from likelihood analysis of the data and the theory multipoles help us determine the optimized $\chi^2$ and the best fit model at the chosen minimum fitting scale. This analysis forms an essential part of the research that we present in this work.  

For all the calculations in our analysis we use $25 \ h^{-1}$Mpc as the minimum fitting scale and $150 \ h^{-1}$Mpc as the maximum fitting scale. The choice of the fitting scales is inspired by \citet{Alam2015a} and follows closely the results presented in there. We use evenly spaced bins of $r$ with a bin size of $5 \ h^{-1}$Mpc. It's instructive to measure the deviation of the data vectors from the model vectors by measuring the $\chi^2/dof$ for each calculation, where $dof$ pertains to the degree of freedom in the analysis. The degree of freedom is obtained from the knowledge of the bin size, minimum ($r_{\texttt{min}}$) and maximum ($r_{\texttt{max}}$) fitting scales of each multipole and the number of comsological parameters ($N_{\rm C}$) that are sampled by \textsc{CosmoMC}. For each instance of data (and theory) we use binned values of $r$ from both the monopole and the quadrupole. Hence, the computation of the $dof$ takes into account the numbers of bins in $r$ in both the monopole and the quadrupole. We use the following equation to compute the degree of freedom: $dof = \left[ \left( r_{\texttt{max}} - r_{\texttt{min}} \right) / \texttt{bin size}\right] \times 2 - N_{\rm C}$.


The obtained values of $\chi^2/dof$ give us an idea of the goodness of the fit of the best fit model with data.

\section{Results} \label{sec:Results}

\begin{table*}
	\centering
	\caption{Statistics of the CLPT-GSRSD fits of MD-P mocks in the three redshift bins. Sampling parameters are the parameters which are sampled in the MCMC algorithm. Derived parameters represent the parameters which are obtained as functions of the sampling parameters. In each bin, we analyze different mocks using the prescription outlined in section~\ref{sec:Analysis} to obtain best fit values and errors for chosen cosmological parameters for each mock. From the statistics obtained for all the mocks, we compute the mean (denoted by $\left\langle \cdot \right\rangle$) and the standard deviations ($S$) of the parameters.}
	\label{tab:MDPResult}
	\begin{tabular}{l|cccccccccccc}
		\hline
		\hline
		Redshift ($z_{\rm eff}$) & \multicolumn{10}{c}{\textbf{Sampling Parameters}} \\
		& $\left\langle \alpha \right\rangle$ & $S_{\alpha}$ & $\left\langle \epsilon \right\rangle$ & $S_{\epsilon}$ & $\left\langle f \right\rangle$ & $S_f$ & $\left\langle F_1 \right\rangle$ & $S_{F_1}$ & $\left\langle F_2 \right\rangle$ & $S_{F_2}$ & $\left\langle \sigma_{\rm FOG} \right\rangle$ & $S_{\sigma_{\rm FOG}}$ \\
		\hline
		0.38 & 0.990 & 0.020 & -0.0046 & 0.0252 & 0.720 & 0.102 & 0.962 & 0.074 & 0.77 & 0.78 & 3.26 & 0.90 \\
		0.51 & 0.993 & 0.017 & -0.0069 & 0.0222 & 0.793 & 0.098 & 1.020 & 0.072 & 0.60 & 0.79 & 3.56 & 0.89 \\
		0.61 & 0.993 & 0.016 & -0.0071 & 0.0220 & 0.833 & 0.099 & 1.112 & 0.069 & 0.49 & 0.84 & 3.87 & 0.98 \\		
		\hline
		& \multicolumn{10}{c}{\textbf{Derived Parameters}} \\
        & & $\left\langle \alpha_{||} \right\rangle$ & $S_{||}$ & $\left\langle \alpha_{\perp} \right\rangle$ & $S_{\perp}$ & $\left\langle f \sigma_8 \right\rangle$ & $S_{f \sigma_8}$ &&&&& \\
		\hline
		0.38 & & 0.984 & 0.058 & 0.995 & 0.028 & 0.472 & 0.067 &&&&& \\
        0.51 & & 0.981 & 0.050 & 1.000 & 0.024 & 0.487 & 0.060 &&&&& \\
        0.61 & & 0.981 & 0.048 & 1.001 & 0.025 & 0.487 & 0.058 &&&&& \\
		\hline		
	\end{tabular}
\end{table*}

In this section we discuss the results from mocks and the BOSS DR12 galaxy data. 

\subsection{Correlation function results on the challenge mocks}
In order to check the correspondence between the different modeling and measurement techniques which have been used in the analysis of the BOSS DR12 combined sample galaxy dataset (including the methods that are combined to obtain the final consensus constrains in \citet{Acacia2016}), the different techniques were tried on large-volume synthetic catalogs in a RSD-fit challenge and compared with each other \citep[][]{Tinker2016}. In addition to checking the agreement of the cosmological information extracted from the different full-shape approaches, this exercise also served to check for possible systematics. We refer the readers \citet{Tinker2016} to see more details of this RSD data challanege. 

Seven different HOD galaxy samples which are built from large-volume $N$-body simulations are investigated in the first part of the challenge. The aforesaid simulations are in tune with $\Lambda$CDM cosmology with moderately different density parameters. The difference between the recovered cosmological parameters and true cosmological parameters were smaller than the statistical error bars for all the methods. The fact that methods based on both the configuration and the momentum spaces show great accuracy and agreement in the constraints that were obtained in the challenge catalogs is very encouraging.

The next part of the challenge involved a set of 84 synthetic catalogs which mimic the North Galactic Coordinate (NGC) part of the DR12 CMASS galaxies (`cut-sky' mocks). These are called the `N series' samples. All of these mocks are obtained using a standard HOD model and are generated from $N$-body simulations which assume the same set of cosmological parameters. As a result, these mocks serve to check for systematic biases in the obtained parameter constraints. In our analysis for the N series mocks, the biases in our recovered parameters were much smaller than the statistical error bars.

\subsection{Results from MD-P mocks}

In Figure~\ref{fig:CorrMatrices}, we present the correlation matrices obtained from the three redshift bins of the MD-P Patchy mocks. In each plot in Figure~\ref{fig:CorrMatrices}, the upper right corner shows the correlation between the bins in the monopole and the quadrupole, the upper left corner shows the correlation between the bins in the quadrupole, the lower right corner represents the correlation between bins in the monopole while the lower left corner illustrates the correlation between the bins in the quadrupole and the monopole.  

%
We independently fit the theory to the correlation function monopoles and quadrupoles for each of the 997 mocks using the covariance matrices computed from the MD-P mocks (for the three redshift bins). The best-fit results are obtained from the comparison of the correlation function multipoles from the mocks with the corresponding multipoles obtained from CLPT-GSRSD theory. The multipoles are binned in evenly spaced values of $s$ with a bin size of $5 \ h^{-1}$Mpc. The tally of data versus theory is done over a fitting range $25 \ h^{-1}$Mpc $ \leq s \leq 150 \ h^{-1}$Mpc. The statistics for the parameters are obtained from the analysis of 997 mocks in each redshift bin. Table~\ref{tab:MDPResult} encapsulates the mean and the standard deviations of the best-fit results of the six parameters that have been sampled by \textsc{CosmoMC}, $viz.$ $\lbrace F_1, F_2, f, \sigma_{\rm FOG}, \alpha, \epsilon \rbrace$. We also catalog the statistics of various derived parameters ($f \sigma_8, b \sigma_8, D_{\rm A}$ and $H$) which are obtained as functions of the sampling parameters in Table~\ref{tab:MDPResult}. 

\begin{figure*}
\begin{multicols}{4}
    \includegraphics[width=\linewidth]{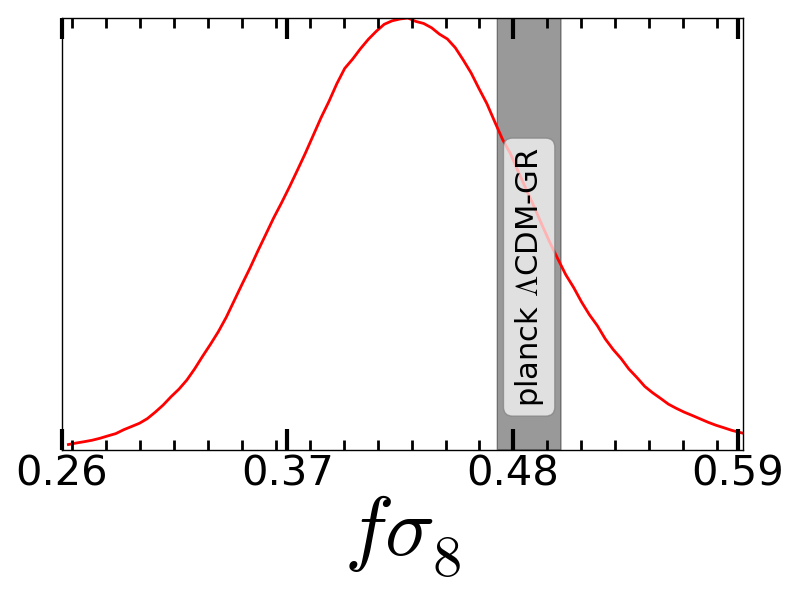}\par
    \includegraphics[width=\linewidth]{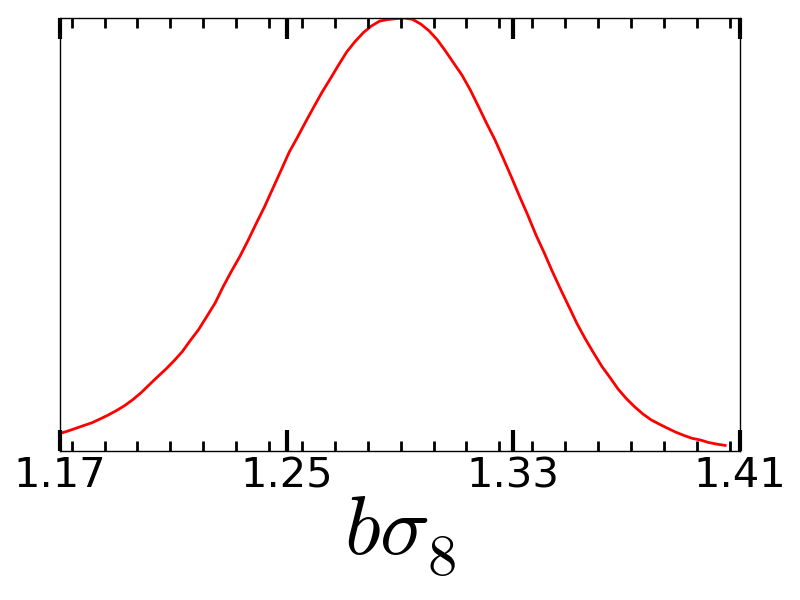}\par
    \includegraphics[width=\linewidth]{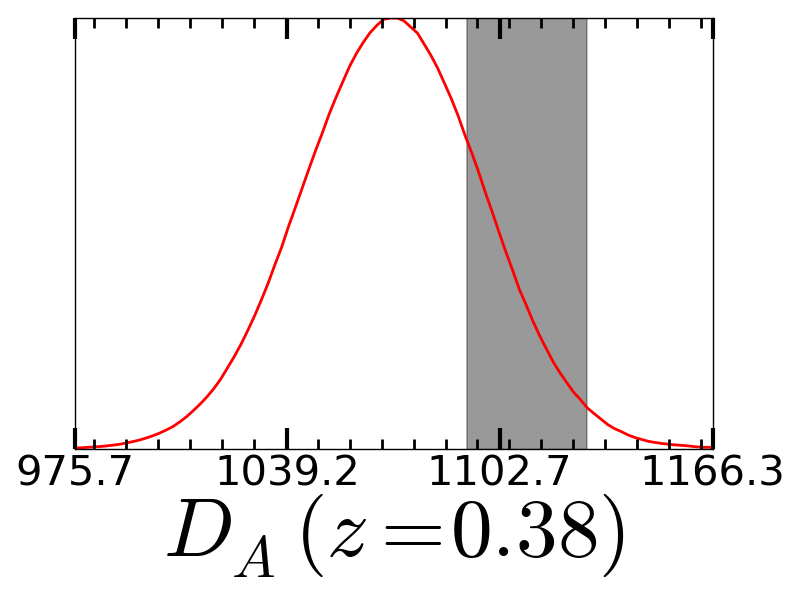}\par
    \includegraphics[width=\linewidth]{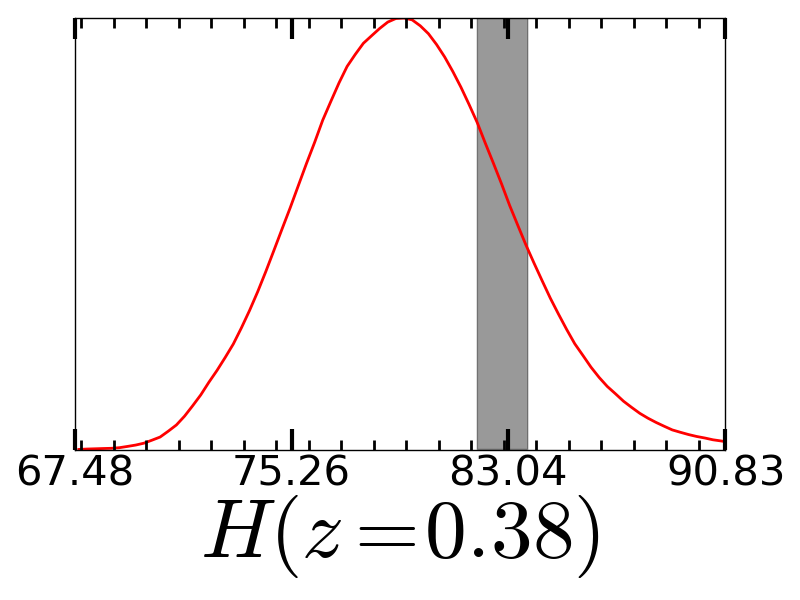}\par    
    \end{multicols}
\begin{multicols}{4}   
    \includegraphics[width=\linewidth]{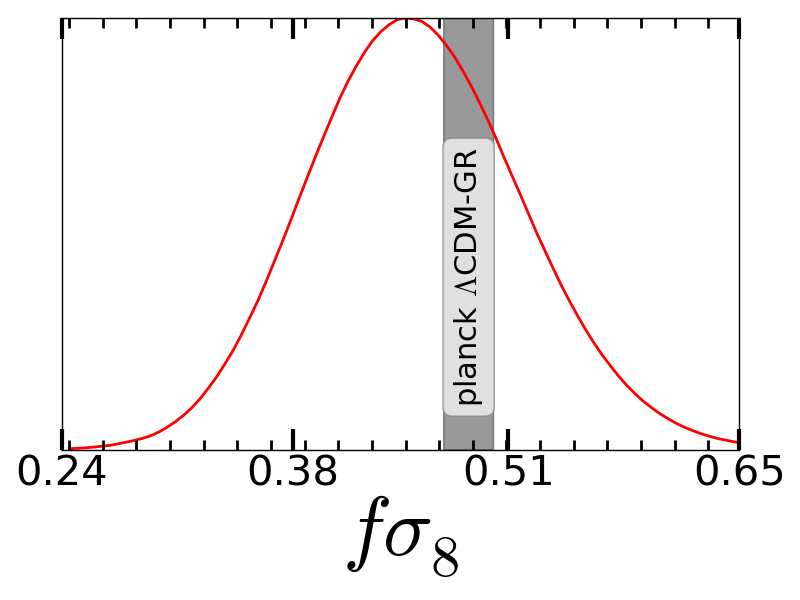}\par
    \includegraphics[width=\linewidth]{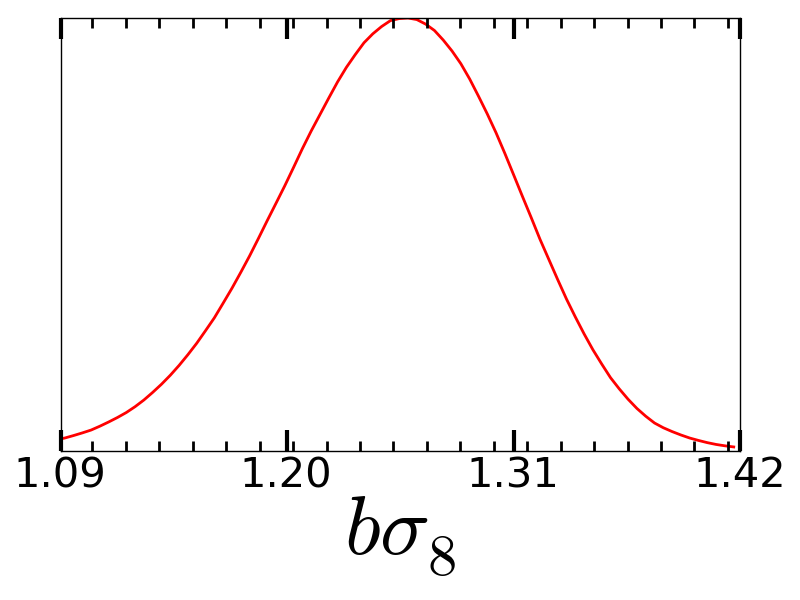}\par
    \includegraphics[width=\linewidth]{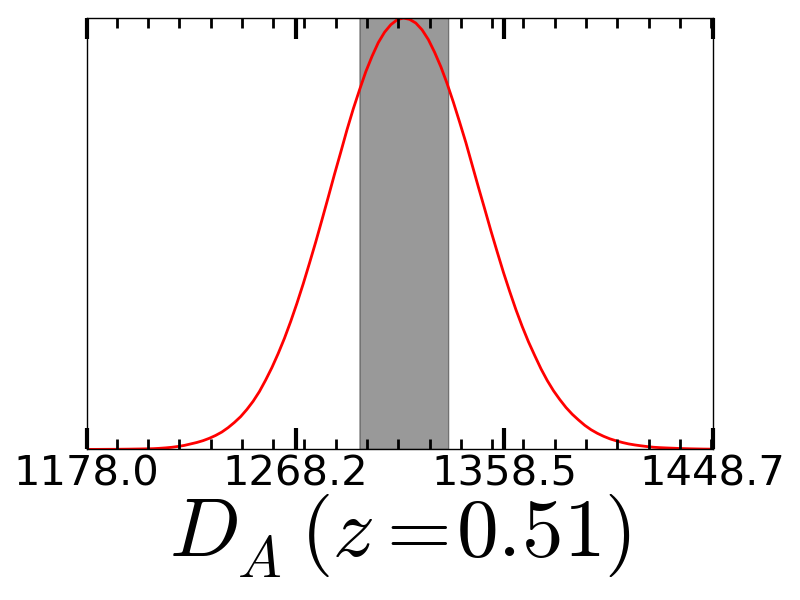}\par
    \includegraphics[width=\linewidth]{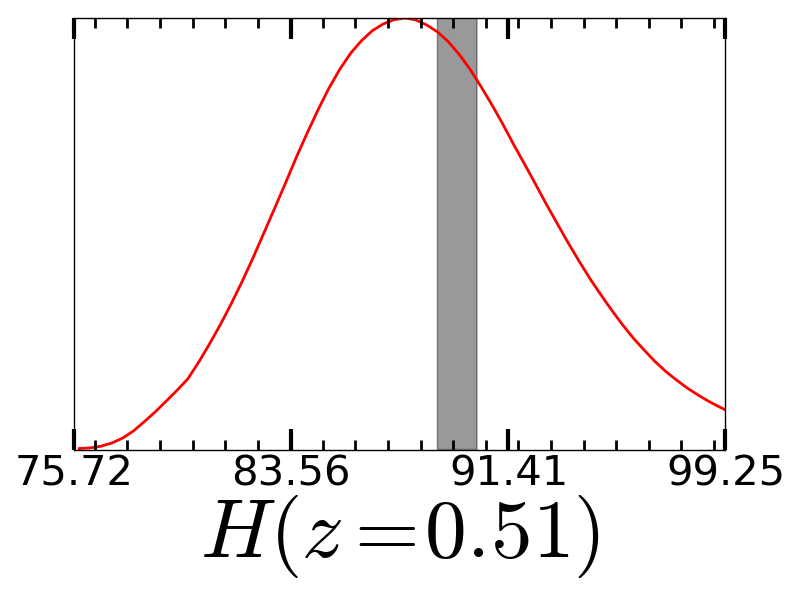}\par 
    \end{multicols}
\begin{multicols}{4}   
    \includegraphics[width=\linewidth]{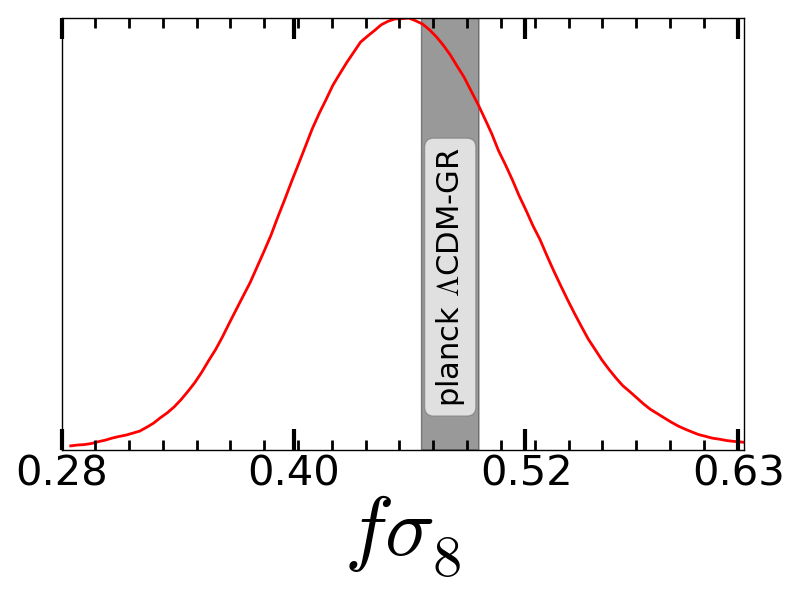}\par
    \includegraphics[width=\linewidth]{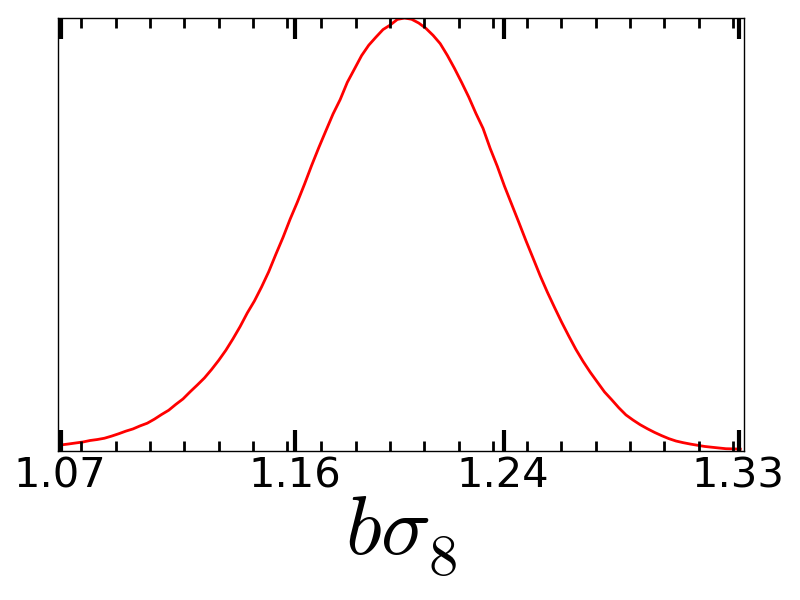}\par
    \includegraphics[width=\linewidth]{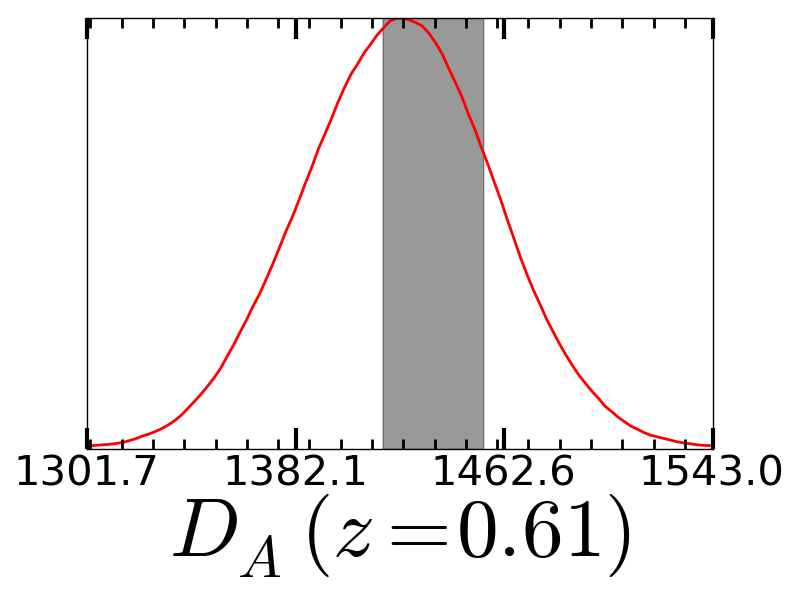}\par
    \includegraphics[width=\linewidth]{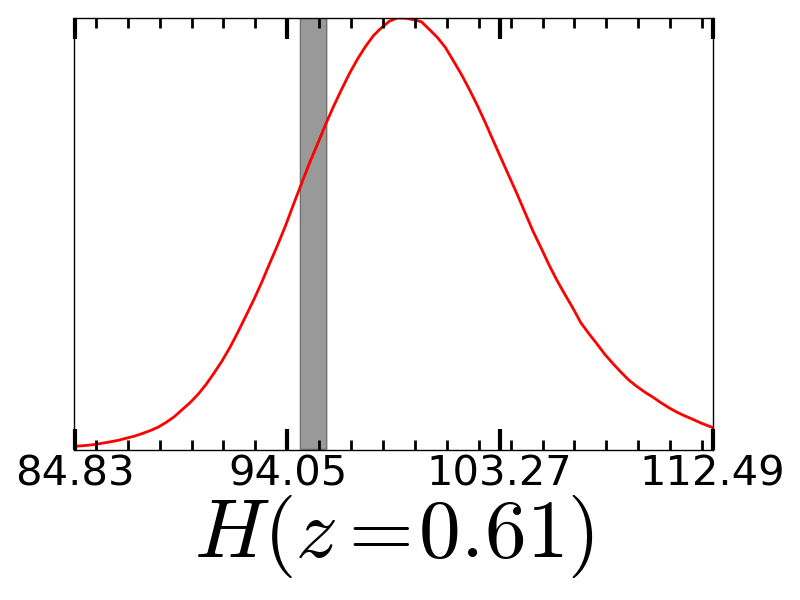}\par 
    \end{multicols}    
    \caption{One dimensional marginalized likelihoods for the parameters $\lbrace f \sigma_8, b \sigma_8, D_{\rm A}, H \rbrace$ for the three redshift bins. The top row depicts results of one dimensional marginalized likelihoods of parameters for the redshift bin with $0.2<z<0.5$, the middle row represents results of one dimensional marginalized likelihoods of parameters for the redshift bin with $0.4<z<0.6$ while the bottom row represents results of one dimensional marginalized likelihoods of parameters for the redshift bin with $0.5<z<0.75$. The grey shaded regions represent $1 \sigma$ spreads of the Planck $\Lambda$CDM predictions of the parameters.}
    \label{fig:Marginzalized_Bin}
\end{figure*}

The values that we get for $f\sigma_8, \alpha_{||}$ and $\alpha_{\perp}$ from our analysis of the MD-P mocks (which are given in Table~\ref{tab:MDPResult}) tally satisfactorily with the corresponding theoretical values of the parameters (which are given in the caption of  Figure~\ref{fig:PatchyHistogram}). Figure~\ref{fig:PatchyHistogram} illustrates the statistical spread of the derived parameters $\alpha_{||}, \alpha_{\perp}$ and $f \sigma_8$ for the three bins in the form of histograms. In Figure~\ref{fig:PatchyHistogram} we show the distribution of the ratios of the differences between the derived parameters $\alpha_{||}, \alpha_{\perp}, f \sigma_8$ and their respective theoretical values $(f \sigma_8)^{th}, \ (\alpha_{||})^{th}, \ (\alpha_{\perp})^{th}$ with their respective standard deviations $S_{f \sigma_8}$, $S_{||}$, $S_{\perp}$. The statistics shown in Figure~\ref{fig:PatchyHistogram} are obtained from the fitting of the multipoles of each of the 997 mocks (in each redshift bin) with multipoles from the theory.

\subsection{Results for BOSS DR12 Galaxy Data} \label{sec:BOSSDR12GalaxyData}

\defcitealias{smith2014}{Paper~I}
\begin{table*}
 \caption{These are the results of the constraints for parameters in our analysis obtained from the BOSS DR12 dataset analysis. In our analysis we have used the fitting range $25 \ h^{-1}$Mpc$ \leq s \leq 150 \ h^{-1}$Mpc with a bin size of $5 \ h^{-1}$Mpc. For each parameter, we report the prior range, mean and $1\sigma$ error. Sampling parameters are parameters which are a part of the parameter space which is sampled by \textsc{CosmoMC}. Derived parameters are parameters which are obtained as functions of the sampling parameters.}
 \label{tab:BOSSDR12Result}
 \begin{tabular}{lcccc}
    \hline
    \hline
    Parameter & prior range & \textbf{\textit{Bin1}} ($z_{\rm eff}=0.38$) & \textbf{\textit{Bin2}} ($z_{\rm eff}=0.51$) & \textbf{\textit{Bin3}} ($z_{\rm eff}=0.61$) \\
    \hline
    \textbf{Sampling Parameters} & & & & \\
    \hline
    $\Omega_b h^2$ & [0.02042 , 0.02372] & $0.02207 \pm 0.00026$ & $0.02206 \pm 0.00026$ & $0.02208 \pm  0.00026$ \\        
    $\Omega_c h^2$ & [0.1041 , 0.1351] & $0.11947 \pm 0.00089$ & $0.11953 \pm 0.00086$ & $0.11963 \pm 0.00086$ \\        
    $n_s$ & [0.9146 , 1.009] & $0.9605 \pm 0.0058$ & $0.9616 \pm 0.0059$ & $0.9621 \pm 0.0058$ \\
    $\ln (10^{10}A_s)$ & [2.670 , 3.535] & $3.076 \pm 0.073$ & $3.078 \pm 0.070$ & $3.089 \pm 0.068$ \\
    $\alpha$ & [0.8000 , 1.350] & $0.991 \pm 0.016$ & $1.006 \pm 0.016$ & $0.980 \pm 0.016$ \\
    $\epsilon$ & [-0.5000 , 0.5000] & $ 0.0275 \pm 0.0183$ & $0.0045 \pm 0.0189$ & $-0.0111 \pm 0.0191$ \\
    $f=d \ln (D)/d \ln (a)$ & [0.1000 , 1.200] & $0.638 \pm 0.080$ & $0.715 \pm 0.090$ & $0.753 \pm 0.088$ \\
    $\sigma_{\rm FOG}$ & [0 , 20] & $3.21 \pm 1.68$ & $3.19 \pm 1.80$ & $2.65 \pm 1.53$ \\
    $F_1$ & [0.5 , 1.5] & $0.91 \pm 0.09$ & $0.98 \pm 0.10$ & $0.98 \pm 0.09$ \\
    $F_2$ & [-5.0 , 5.0] & $-0.47 \pm 0.31$ & $-0.2 \pm 1.48$ & $-0.51 \pm 0.32$ \\
    \hline
    \textbf{Derived Parameters} & & & & \\
    \hline
    $f \sigma_8$ & $\texttt{x}$ & $0.430 \pm 0.054$ & $0.452 \pm 0.057$ & $0.457 \pm 0.052$ \\
    $b \sigma_8$ & $\texttt{x}$ & $1.283 \pm 0.039$ & $1.249 \pm 0.051$ & $1.198 \pm 0.036$ \\
    $D_{\rm A}$ & $\texttt{x}$ & $1069.6 \pm 24.0$ & $1314.5 \pm 27.4$ & $1420.6 \pm 33.4$ \\
    $H$ & $\texttt{x}$ & $79.3 \pm 3.2$ & $88.4 \pm 4.1$ & $99.5 \pm 4.4$ \\
    \hline 
 \end{tabular}
\end{table*}

We follow the steps outlined in section~\ref{sec:Analysis} to obtain optimized values of cosmological parameters for the three redshift bins of the BOSS DR12 galaxy dataset. At this stage, it is worthwhile to note that in our analysis of the BOSS DR12 galaxy data we use the covariance matrices obtained from the MD-P mocks. In Table~\ref{tab:BOSSDR12Result}, we list the parameters used in our analysis and the best-fit results returned by \textsc{CosmoMC} for the parameters for the three galaxy bins ($z_{\rm eff}=0.38, \ 0.51$ and $0.61$). Our results for the parameters $f \sigma_8$, $b \sigma_8$, $D_{\rm A}$ and $H$ for bins 1, 2 and 3 agree with reports of results of parameters from efforts by other groups \citep[][]{Acacia2016}.

Figure~\ref{fig:CombBin} illustrates the monopole and the quadrupole for BOSS DR12 data and the best fit models of the data for the three bins. The black dots in the plots of Figure~\ref{fig:CombBin} represent the measured correlation functions for BOSS DR12 galaxy sample. The diagonal elements of the covariance matrices corresponding to mocks in the three redshift bins give us the error bars for our observations in the three bins. As indicated in the plots, the red and the blue lines denote the best fits of monopole and quadrupole respectively using the fitting range $25 \ h^{-1}$Mpc$ \leq s \leq 150 \ h^{-1}$Mpc with a bin size of $5 \ h^{-1}$Mpc. Our results show that the means of the best-fit models for the monopole and the quadrupole tally with the multipoles from the data very well. This is evidenced by the values of $\chi^2/dof$ ($53/44, 52/44, 41/44$) that we obtain for the comparison of multipoles of data and theory of the three redshift bins. The values of $\chi^2/dof$ are as per expectations. Figure~\ref{fig:Marginzalized_Bin} shows the results of the marginalized likelihood of the derived parameters $\lbrace f \sigma_8, b \sigma_8, D_{\rm A}, H \rbrace$.

\begin{figure}
    \includegraphics[width=\columnwidth]{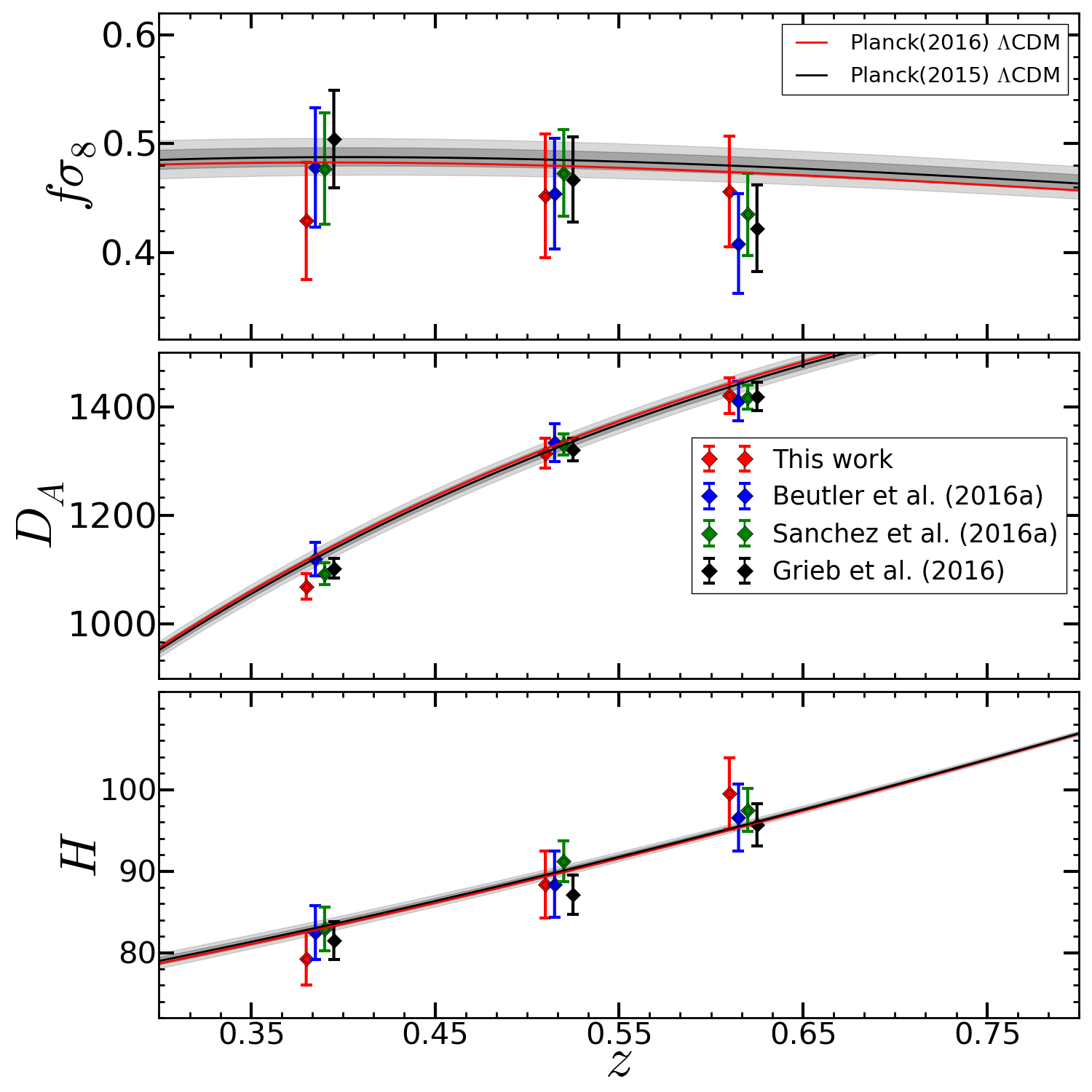}
    \caption{Here we compare our results of $f\sigma_8(z), D_{\rm A}(z)$ and $H(z)$ with the predictions of Planck $\Lambda$CDM and with the results of \citet{Beutler2016a}, \citet{Grieb2016} and \citet{Sanchez2016a}. The dark and the light shaded regions represent the $1 \sigma$ and the $2 \sigma$ spreads of the Planck $\Lambda$CDM (TTTEEE+lowP) predictions of $f\sigma_8, D_A$ and $H$. The solid black line shows the \citet{Planck2015} predictions for the variation of $f\sigma_8$ as a function of $z$ while the solid red line shows the predictions of \citet{Planck2016} for the variation of $f\sigma_8$ with respect to $z$.}
    \label{fig:Comb}
\end{figure}

\begin{figure}
\begin{multicols}{2}
    \includegraphics[width=\linewidth]{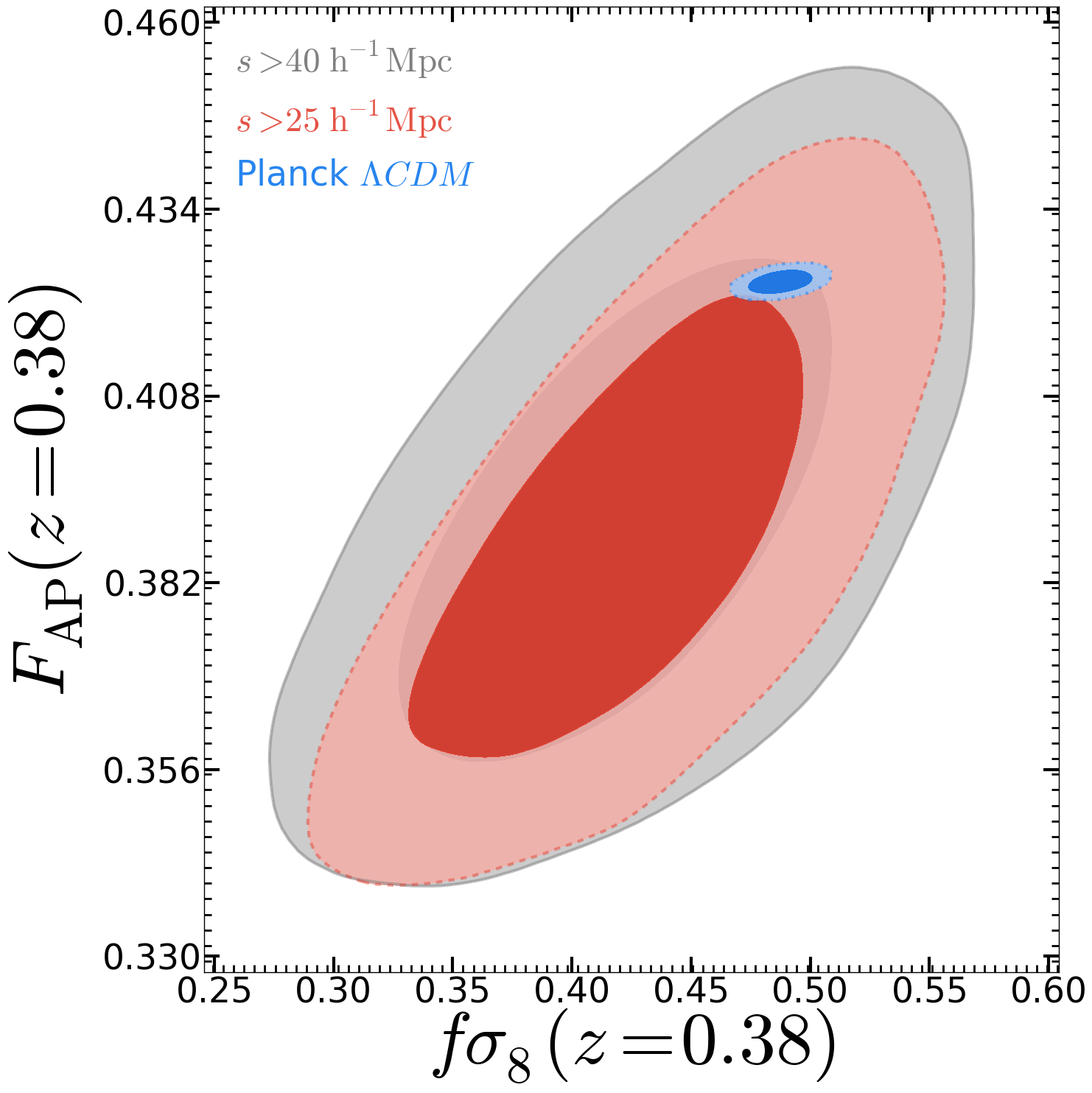}\par
    \includegraphics[width=\linewidth]{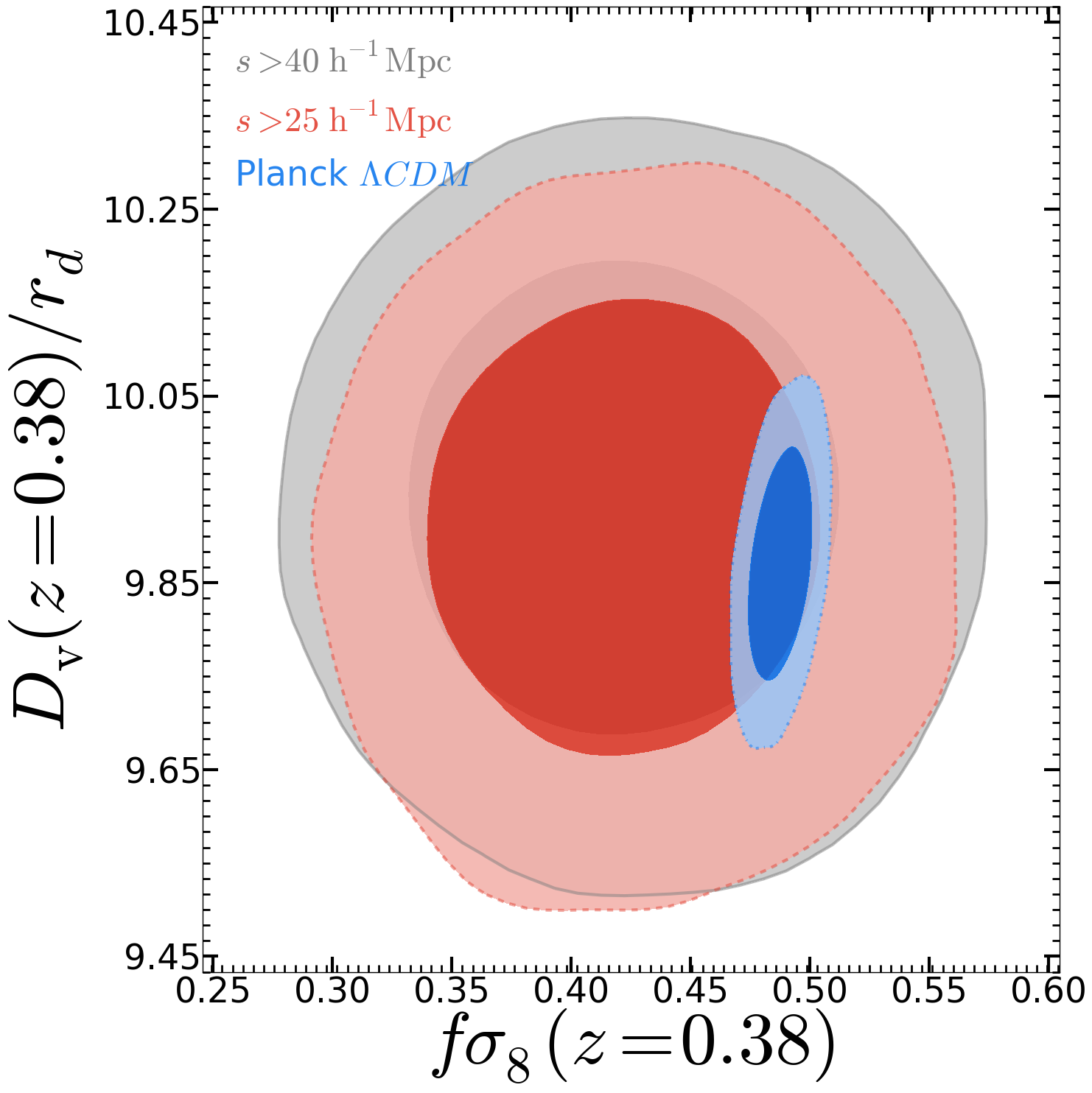}\par        
    \end{multicols}
\begin{multicols}{2}
    \includegraphics[width=\linewidth]{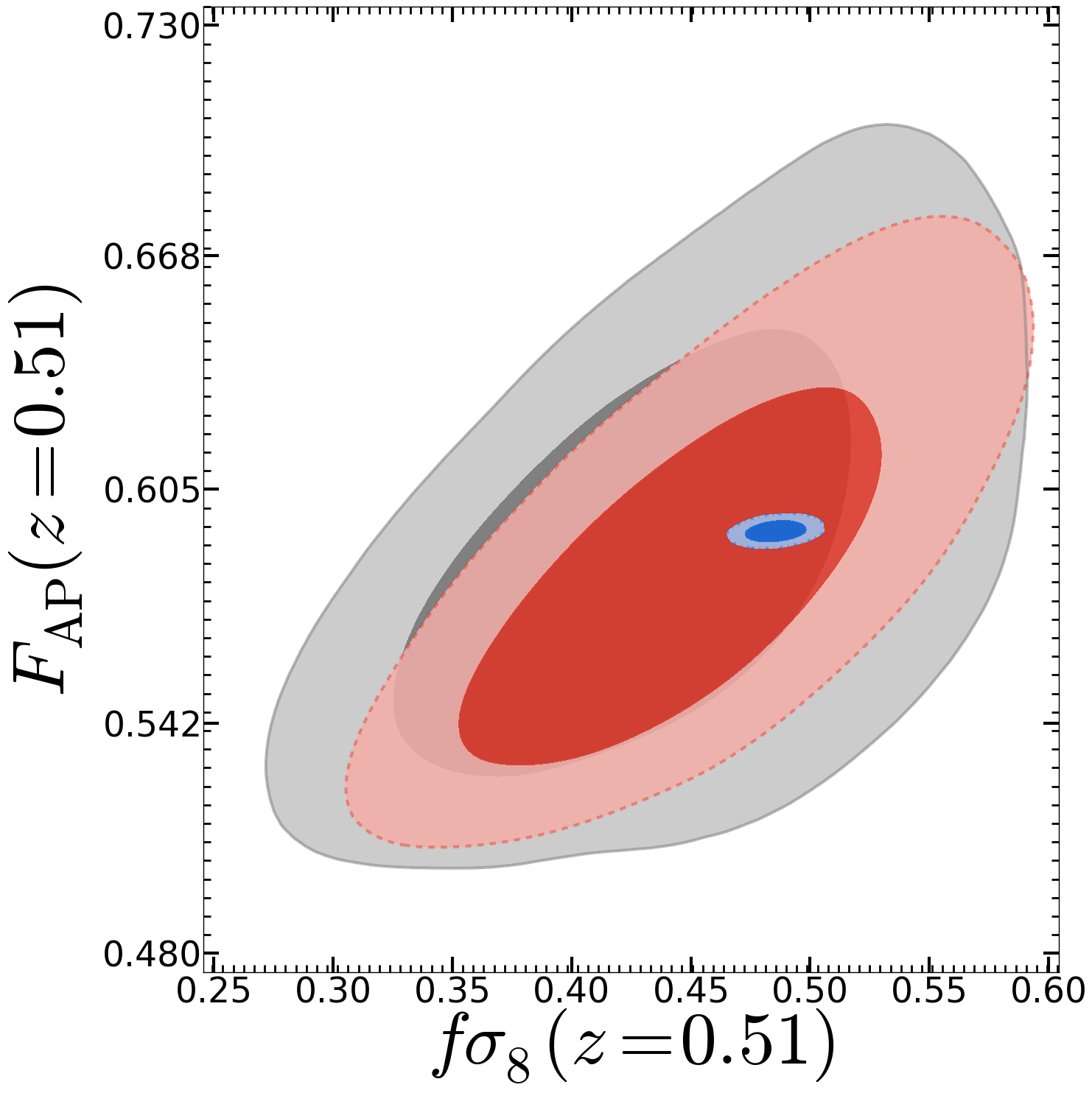}\par       
    \includegraphics[width=\linewidth]{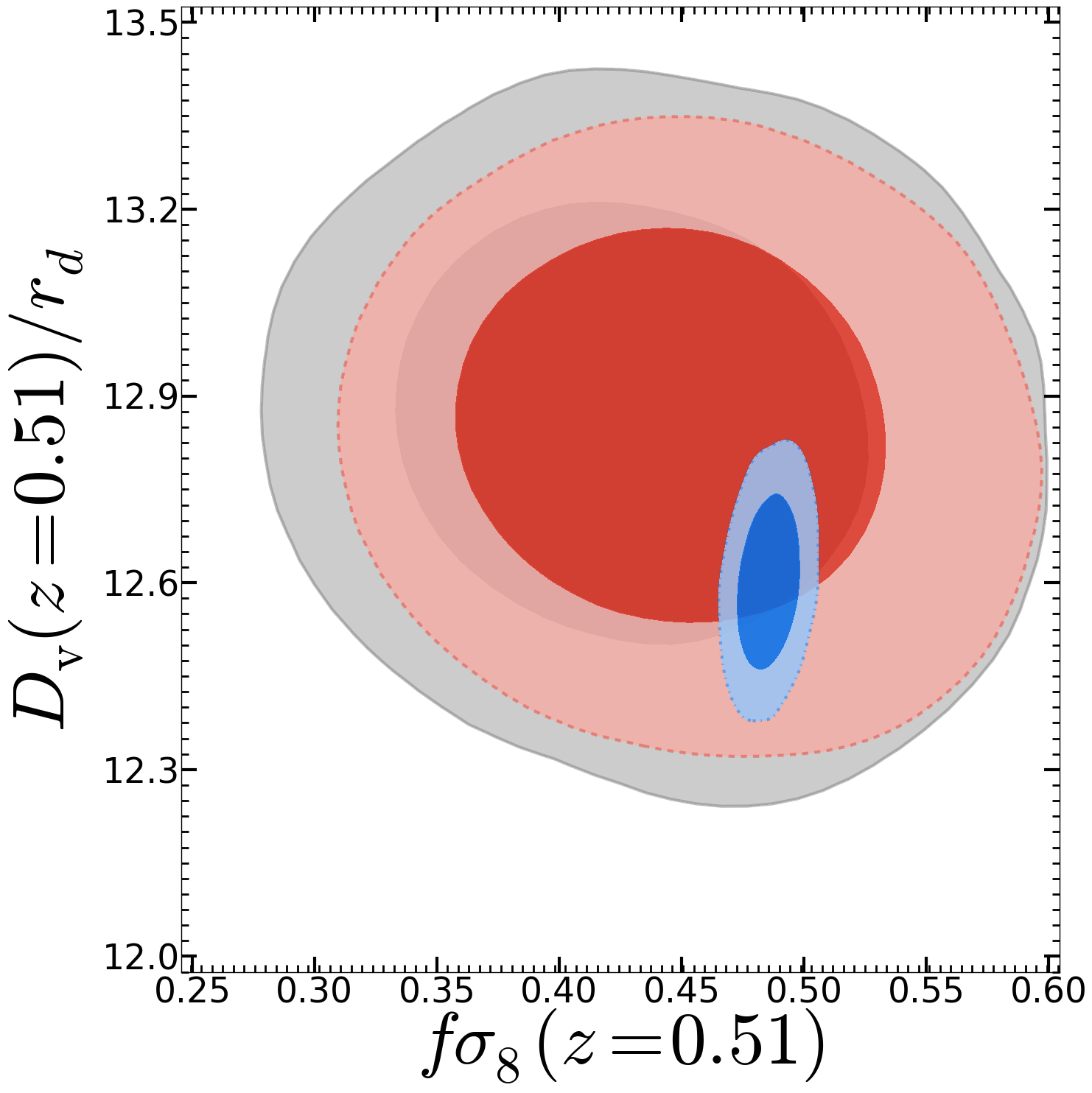}\par
    \end{multicols}
\begin{multicols}{2} 
    \includegraphics[width=\linewidth]{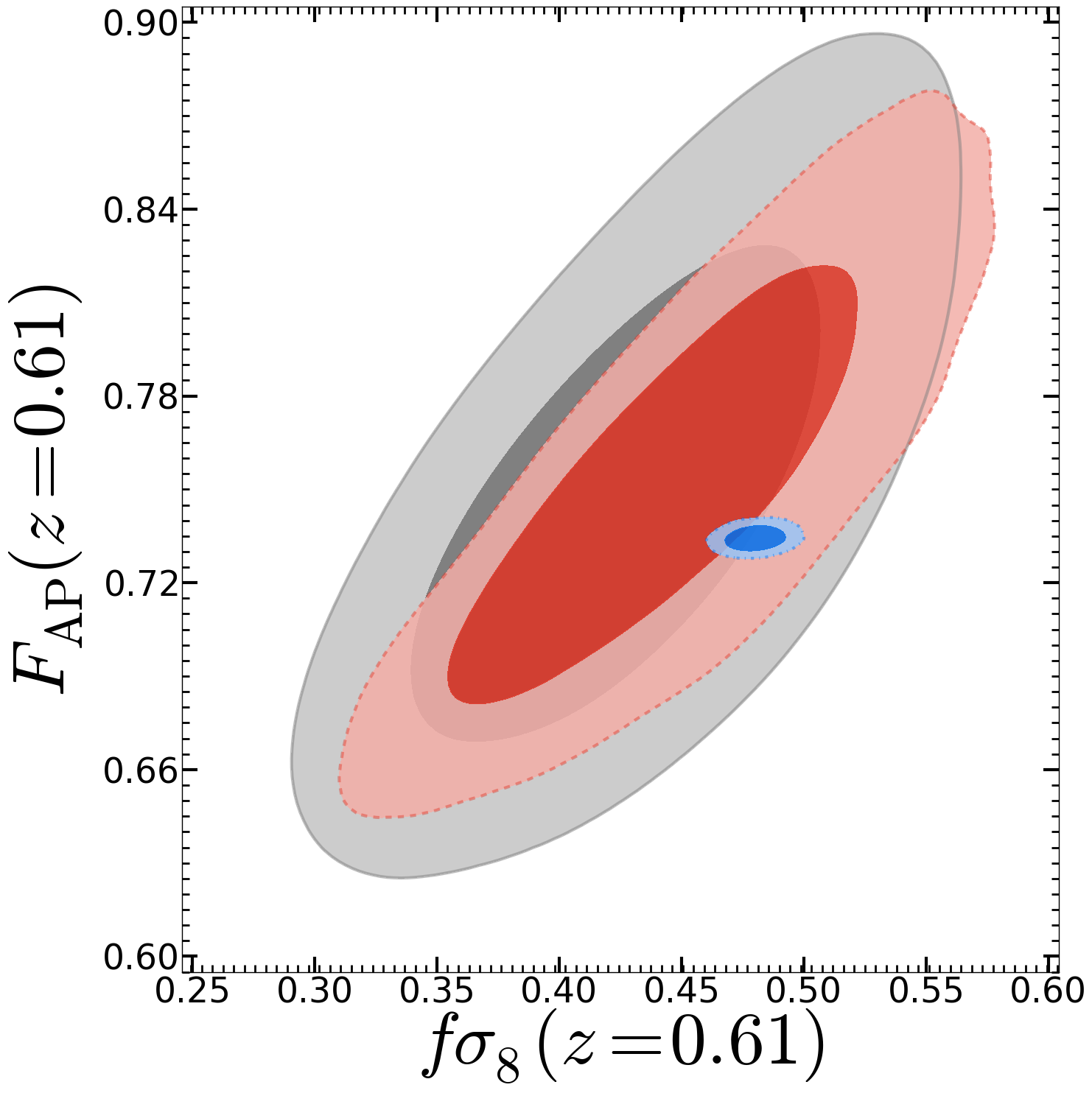}\par   
    \includegraphics[width=\linewidth]{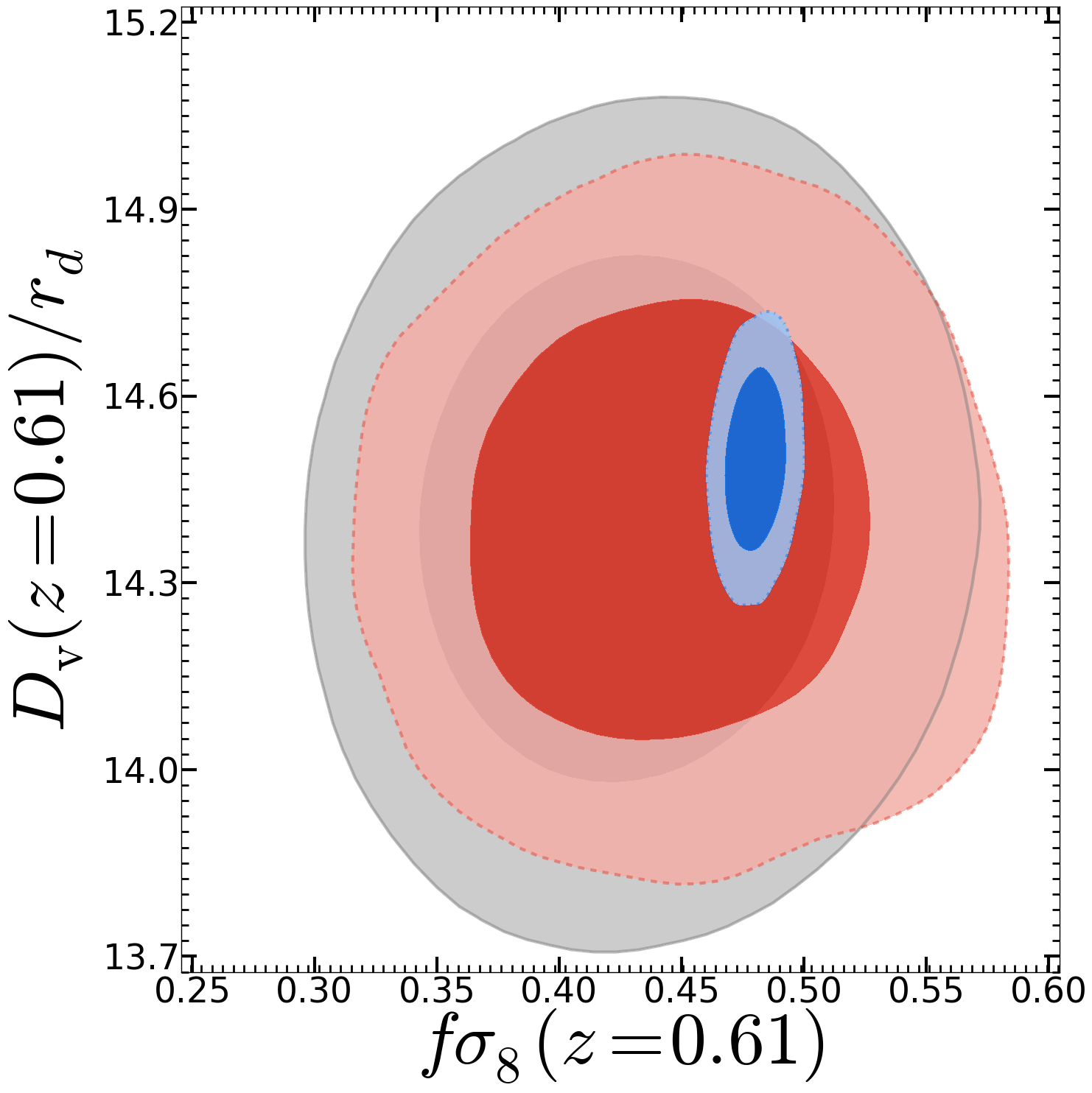}\par 
    \end{multicols}    
    \caption{In this figure we show $1 \sigma$ ($68 \%$) and $2 \sigma$ ($95 \%$) confidence intervals for $F_{\rm AP}$ and $f \sigma_8$ on the left and $D_{\rm v} / r_{\rm d}$ and $f \sigma_8$ on the right.  The grey contours show constraints from larger scales ($s > 40 \ h^{-1}$Mpc) while the red contours depict constraints from all scales ($s > 25 \ h^{-1}$Mpc) from our RSD analysis. The blue contour shows the constraints from Planck 2015 results. The top, middle and the bottom rows are from the three redshift bins, $viz.$ $z_{\rm eff}=0.38, \  0.51$ and $0.61$ respectively.  All estimates are mutually consistent.}
    \label{fig: CosmoMcLikelihood}
\end{figure}

\section{Discussion} \label{sec:Discussion}

Our measurements of $f \sigma_8$, $D_{\rm A}$ and $H$ using CLPT-GSRSD are consistent with the predictions of the $\Lambda$CDM for all the three redshift bins. Furthermore, our results for $f \sigma_8$, $D_{\rm A}$ and $H$ obtained from CLPT-GSRSD agree very well with measurements of the same parameters obtained from other approaches which do investigations based on full-shape analyses of SDSS-III BOSS DR12 combined sample \citep[][]{Beutler2016a, Grieb2016, Sanchez2016a}. At the same time, the theoretical models used by the four analyses are very different. A comparison of the different results is presented in Figure~\ref{fig:Comb}. The black and the red lines in the plot of $ f \sigma_8$ vs $z$ in Figure~\ref{fig:Comb} show the predictions of \citet{Planck2015} and \citet{Planck2016} respectively. The difference between the predictions of Planck 2015 and Planck 2016 is not more than $1/2$ $\sigma$. The correspondence of results of $f \sigma_8$ at multiple redshifts which are obtained from different theories can be considered as a useful probe of the theory of gravity. It also holds the promise of letting us place model independent constraints on other models of gravity. 

One of the challenges in RSD analysis is to use the smaller scales as they have higher signal to noise by virtue of sampling large numbers of two point modes. But, perturbation theory based models find it difficult to describe measurements at smaller scales due to non-linear clustering. The model used in our analysis has been validated using various approximate mocks and N-body mocks. We fit to scales upto 25 $h^{-1}$Mpc in our final analysis. In order to understand the contribution from quasi-linear scales and to further look for biases in our analysis we have also run our analysis using a linear scale of $s>40 h^{-1}$Mpc. Figure~\ref{fig: CosmoMcLikelihood} shows the comparison between results obtained from fitting only linear scale and results obtained while including the quasi linear scales. In Figure~\ref{fig: CosmoMcLikelihood},  we show $1 \sigma$ ($68 \%$) and $2 \sigma$ ($95  \%$) confidence intervals for $F_{\rm AP}$ and $f \sigma_8$ on the left and $D_{\rm v} / r_{\rm d}$ and $f \sigma_8$ on the right.  The grey contours show constraints from larger scales ($s > 40 \ h^{-1}$Mpc) while the red contours depict constraints from all scales ($s > 25 \ h^{-1}$Mpc) from our RSD analysis. The blue contour shows the constraints from Planck 2015 results. The top, middle and the bottom rows are from the three redshift bins, $viz.$ $z_{\rm eff}=0.38, \ 0.51$ and $0.61$ respectively.  The inclusion of the quasi-linear scale improves the constraints without introducing any statistically significant shifts in the measurements.  The improvement in $F_{\rm AP}$ and $f\sigma_8$ is larger compared to $D_{\rm V} / r_{\rm d}$ because most of the information in $D_{\rm V} / r_{\rm d}$ is contained in the BAO peak.

\begin{figure}
    \includegraphics[width=\columnwidth]{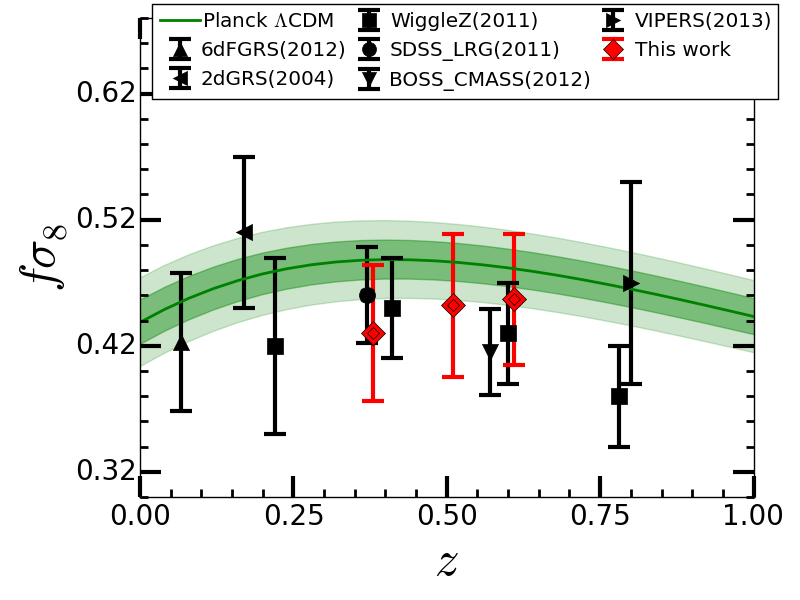}
    \caption{Here, we plot measurements of $f\sigma_8$ from different surveys and research studies. The included surveys and studies report measurements of $f\sigma_8$ over a redshift range of $0.06 < z < 0.80$. We have represented the $1 \sigma$ and the $2 \sigma$ spreads of the Planck $\Lambda$CDM prediction for the evolution of $f\sigma_8$ with redshift in the dark green and the light green shaded regions respectively.}
    \label{fig:WorldCompilation}
\end{figure}

Figure~\ref{fig:WorldCompilation} presents a compilation of $f\sigma_8$ measurements at different redshifts from different surveys and research studies. We expect our results to provide a robust test of the underlying theory of gravity at large distance scales.

We decided not to push for measurements of the linear growth rate of structure at fitting scales smaller than the minimum fitting scale that we have chosen here because of the lack of reliability in the behavior of the model at small distance scales. From the perspective of a comprehensive RSD analysis it will be invaluable to gain knowledge of estimates of cosmological parameters at distance scales smaller than $20 \ h^{-1}$Mpc. However, an analysis of smaller cosmological scales with presently available theoretical resources presents a significant challenge. The theoretical models that are currently available to us are unable to model small distance scales. It is worthwhile to investigate if this inability is due to the presence of the non-linear Finger of God effect. Are the contributions from non-linear clustering (which are different from the Fingers of God effect) modeled accurately? These are questions that we wish to seek answers to. We would like to investigate the efficacy of the Gaussian Streaming Model in explaining non-linear clustering at small scales. As an outlook for the future, we also plan to explore the feasibility of designing and using new estimators to probe scales smaller than $20 \ h^{-1}$Mpc and also test the effectiveness of CLPT-GSRSD at such scales.  

\section{Summary} \label{sec:Summary}
We have used CLPT-GSRSD to measure cosmological parameters including the linear growth rate of structure $f$ from the SDSS III BOSS DR12 combined galaxy sample. The BOSS DR12 combined galaxy dataset includes over a million massive galaxies encompassing a redshift range $0.2<z<0.75$. We divide this sample to three partially overlapping redshift bins with effective redshifts of 0.38, 0.51 and 0.61 and we work with multipole moments of two-point galaxy correlation functions in these three redshift bins.  We use the measured and best fit multipole moments to place constraints on cosmological parameters including the linear growth rate of structure in the Universe. The fitting scale that we choose in this work is dictated by the performance, reliability and considerations of error of MD-P mocks and our theoretical model at small scales. Our measurements of the growth rate of structure, $f \sigma_8(z)$, angular diameter distance $D_A(z)$ and the Hubble expansion rate, $H(z)$ are in agreement with the results for the same parameters obtained by different groups \citep[][]{Beutler2016a, Grieb2016, Sanchez2016a}. Furthermore, our results are combined with other BAO \citep[][]{Beutler2016b, Ross2016, Vargas2016} and full-shape methods in a set of final consensus constraints in \citet{Acacia2016}. Our results are in consonance with the predictions of the Planck $\Lambda$CDM model. We expect the results of our work to shed more light on the evolution of the linear growth rate of structure and contribute towards lifting the ambiguity in the choice between dark energy and modified theories of gravity. The measurements we report in this work can contribute to constrain cosmological parameters in different models of gravity. Through our work, we also provoke questions of whether it is possible to model non-linearities at small distance scales in the Universe. 

\section*{Acknowledgements}
We acknowledge useful comments and suggestions from S\'{e}bastien Fromenteau, Alex Geringer-Sameth, Arun Kannawadi, Elena Giusarma  and Anthony R. Pullen. This work was supported by NASA 12-EUCLID11-0004 and NSF AST1412966. Funding for SDSS-III has been provided by the Alfred P. Sloan Foundation, the Participating Institutions, the National Science Foundation, and the U.S. Department of Energy Office of Science. The SDSS-III web site is \href{http://www.sdss3.org/}{http://www.sdss3.org/}.

SDSS-III is managed by the Astrophysical Research Consortium for the Participating Institutions of the SDSS-III Collaboration including the University of Arizona, the Brazilian Participation Group, Brookhaven National Laboratory, Carnegie Mellon University, University of Florida, the French Participation Group, the German Participation Group, Harvard University, the Instituto de Astrofisica de Canarias, the Michigan State/Notre Dame/JINA Participation Group, Johns Hopkins University, Lawrence Berkeley National Laboratory, Max Planck Institute for Astrophysics, Max Planck Institute for Extraterrestrial Physics, New Mexico State University, New York University, Ohio State University, Pennsylvania State University, University of Portsmouth, Princeton University, the Spanish Participation Group, University of Tokyo, University of Utah, Vanderbilt University, University of Virginia, University of Washington, and Yale University.












\bsp	
\label{lastpage}
\end{document}